\documentclass[aps,showpacs,epsf,dvips]{revtex4}
\usepackage{feynmp}
\usepackage{amssymb}
\usepackage{epsfig}
\usepackage{graphicx}
\usepackage{subfigure}
\usepackage{color}

\begin{document}

\title{Non-Fermi liquid behavior due to U(1) gauge field in two dimensions}
\author{Jing-Rong Wang and Guo-Zhu Liu}
\address{\small {\it Department of Modern Physics, University of Science and
Technology of China, Hefei, Anhui, 230026, P.R. China}}

\begin{abstract}
We study the damping rate of massless Dirac fermions due to the U(1)
gauge field in (2+1)-dimensional quantum electrodynamics. In the
absence of a Maxwell term for the gauge field, the fermion damping
rate $\mathrm{Im}\Sigma(\omega,T)$ is found to diverge in both
perturbative and self-consistent results. In the presence of a
Maxwell term, there is still divergence in the perturbative results
for $\mathrm{Im}\Sigma(\omega,T)$. Once the Maxwell term is included
into the self-consistent equations for fermion self-energy and
vacuum polarization functions, the fermion damping rate is free of
divergence and exhibits non-Fermi liquid behavior:
$\mathrm{Im}\Sigma(\omega,T) \propto
\mathrm{max}(\sqrt{\omega},\sqrt{T})$.
\end{abstract}

\maketitle

%%%%%%%%%%%%%%%%%%%%%%%%%%%%%Main Body%%%%%%%%%%%%%%%%%%%%%%%%%%%%%%%%%%%%%

\section{Introduction\label{Sec:Introduction}}

In a normal Fermi liquid, the Landau quasiparticles are well-defined
in the low-energy regime since their damping rate vanishes rapidly
as $\mathrm{Im}\Sigma(\omega,T) \propto
\mathrm{max}(\omega^{2},T^{2})$ upon approaching the Fermi surface.
Such rapidly diminishing damping rate is guaranteed by the Pauli
exclusion principle and can be naturally understood by the fact that
most of the states into which the fermions on Fermi surface would be
scattered are already occupied by other fermions. Indeed, the states
on the Fermi surface have an infinite lifetime and the low-lying
quasiparticles can be safely considered as being nearly independent
even when the interaction is not weak. Generally, one condition for
the stability of Fermi liquid is the absence of singular or
long-range interaction \cite{Varma2002}. In normal metals, although
the bare Coulomb interaction is long-ranged, it becomes short-ranged
after including the dynamical screening effect. Therefore, the Fermi
liquid theory can provide an excellent description for the electron
liquid in metals.

Unlike Coulomb interaction, the long-range gauge interaction usually
can not be fully screened in the absence of gauge symmetry breaking
(it becomes short-ranged in a superconductor via Anderson-Higgs
mechanism). It is thus possible that the unscreened gauge
interaction would generate behaviors those are beyond the scope of
Fermi liquid theory. In the context of ordinary metals, Holstein
\emph{et} \emph{al.} first pointed out that non-Fermi liquid
behavior arises from the coupling of electrons with the unscreened
electromagnetic field \cite{Holstein}. This prominent result began
the adventure for seeking various non-Fermi liquid behaviors in
several different gauge theories \cite{Leereview, Affleck, Kim97,
Kim99, Rantner, Hermele, Kaul, Franz, Herbut, Ran07, Reizer, Ioffe,
Lee92, Nayak} in the subsequent three decades.

The (2+1)-dimensional quantum electrodynamics (QED$_{3}$) of
massless Dirac fermions is an interesting model that has been widely
studied in both high energy physics \cite{Appelquist88, Maris95} and
condensed matter physics \cite{Leereview, Affleck, Kim97, Kim99,
Rantner, Hermele, Kaul, Franz, Herbut, Ran07}. The gauge field is
strongly interacting with Dirac fermions, giving rise to rather
unusual behaviors. As shown by Appelquist \emph{et} \emph{al.}, the
massless Dirac fermions can acquire a finite mass via the mechanism
of dynamical chiral symmetry breaking \cite{Appelquist88}. In the
context of condensed matter physics, this phenomenon is usually
identified as the formation of long-range antiferromagnetism in
two-dimensional  quantum Heisenberg antiferromagnet \cite{Kim99}. In
the phase with chiral symmetry unbroken, the Dirac fermions are
massless and the gauge field is strongly fluctuating but stays in
the deconfined phase \cite{Maris95}. Now this field theory can
describe the physics of a U(1) spin liquid, which is a novel state
of matter without any spontaneous symmetry breaking. It is important
to emphasize that the absence of symmetry breaking does not mean the
absence of interesting physics. In fact, the U(1) spin liquid
manifests a great deal of unusual physical properties and has been
used to understand several correlated electron systems, including
high temperature copper-oxide superconductors \cite{Leereview,
Affleck, Kim97, Kim99, Rantner, Hermele, Kaul} and some spin-1/2
Kagome systems \cite{Ran07}. One of the most remarkable features of
the gauge interaction is its ability to produce non-Fermi liquid
behaviors. Owing to the chirality and linear spectrum of Dirac
fermions, this interacting field theory is expected to display
distinct behaviors compared with the much studied non-relativistic
gauge systems \cite{Reizer, Ioffe, Lee92, Nayak}.

In this paper, we study the possible non-Fermi liquid behavior by
computing the damping rate $\mathrm{Im}\Sigma(\omega,T)$ of massless
Dirac fermion within the QED$_{3}$ theory. Generically, there are
two kinds of QED$_{3}$ theories, depending on whether the theory has
an explicit Maxwell term or not. First of all, as a well-defined
quantum gauge field theory, it contains the Maxwell term $\propto
F_{\mu\nu}F^{\mu\nu}$ explicitly in the action. If the theory is
constructed by considering the quantum phase fluctuations in
underdoped high temperature superconductors, then there is also an
explicit Maxwell term in the Lagrangian\cite{Franz, Herbut}. On the
contrary, when the effective QED$_{3}$ theory is obtained by the
slave-particle treatment of \emph{t}-\emph{J} model, there is no
Maxwell term in the Lagrangian and the gauge field has its own
dynamics only after integrating out the matter fields
\cite{Leereview, Affleck, Kim97, Kim99, Rantner}. Here, we consider
both of these two kinds QED$_{3}$ and show that they behave
differently.

We first consider the QED$_{3}$ theory without Maxwell term. In the
Coulomb gauge, the temporal and spatial components of U(1) gauge
field are decoupled. We calculate the longitudinal and transverse
fermion damping rates at both zero temperature, $T = 0$, and finite
temperature. By straightforward perturbation computation, we find
that the damping rate is always divergent, either in the
longitudinal contribution or in the transverse contribution. The
appearance of divergences indicates the insufficiency of ordinary
perturbative expansion in treating systems with singular
interaction. Moreover, divergence still exists in the
self-consistently coupled equations for fermion damping rate and
gauge boson propagator.

We then consider the case with explicit Maxwell term for the gauge
field. At the perturbative level, the fermion self-energy function
diverges even in the presence of such term. After studying the
self-consistent equations for fermion damping rate and gauge boson
propagator, we found that the fermion damping rate is free of
divergence and given by $\mathrm{Im}\Sigma(\omega,T) \propto
\mathrm{max}(\sqrt{\omega},\sqrt{T})$, which is a non-Fermi liquid
behavior.

We also discuss the damping rate of Dirac fermions due to the
long-range Coulomb interaction. Although Coulomb interaction may be
identified as the non-relativistic counterpart of the U(1) gauge
interaction, they lead to different properties of Dirac fermions.
Marginal Fermi liquid behavior is found by both perturbative and
self-consistent approaches.

The damping rate of massless Dirac fermion is studied in QED$_{3}$
without Maxwell term in Sec. 2 and with Maxwell term in Sec. 3. In
Sec. 4, we consider the Coulomb interaction and discuss why
divergence appears in gauge theory, but not in such system. We end
the paper with a summary and a brief discussion.

\section{Fermion damping rate in QED$_{3}$ without Maxwell term\label{sec:NoMaxwellTerm}}

The Lagrangian of QED$_{3}$ has the form
\begin{equation}
\mathcal{L} = \sum_{i=1}^{N}\Psi_{i}^\dag\left(\partial_{\tau} - ia_{0} - i\mathbf{\sigma}\cdot
\left(\mathbf{\partial} -i\mathbf{a}\right)\right)\Psi_{i} - \frac{1}{4}F_{\mu\nu}F^{\mu\nu}.
\end{equation}
In principle, the fermion field $\Psi$ can be expressed in
four-component or two-component representation. Since we only
consider the chiral symmetric phase, we adopt the two-component
representation of spinor field. The Dirac fermion flavor is taken to
be a general $N$ in order to perform $1/N$ expansion. In this paper,
we discuss only non-compact QED$_{3}$ and hence there are no
instantons. Note that the coupling between massless Dirac fermion
and gauge field respects the Lorentz invariance, so there is no
singular velocity renormalization \cite{Kim97}. When applied to high
temperature superconductor, there is indeed an velocity anisotropy.
However, this anisotropy turns out to be irrelevant, thus restoring
the Lorentz invariance \cite{Vafek}. In this paper, the fermion
velocities are simply taken to be unity.

As mentioned in the Introduction, we will first consider QED$_{3}$
theory without Maxwell term for gauge field. This is the effective
low-energy theory of \emph{t}-\emph{J} model, obtained by using the
slave-particle treatment. Now the $\propto F_{\mu\nu}F^{\mu\nu}$
term is simply dropped from the Lagrangian. The gauge field
appearing in such model has its own dynamics only after integrating
out the fermion fields, as well as other possible matter fields.

The Matsubara propagator of massless Dirac fermion is
\begin{equation}
G_{0}\left(\omega_{n},\mathbf{k}\right) =
\frac{1}{i\omega_{n}-\mathbf{\sigma}\cdot\mathbf{k}},
\end{equation}
where $\omega_{n}=(2n+1)\pi T$ with $n$ being integers. After
analytic continuation, the retarded propagator reads
\begin{equation}
G_{0}(\omega,\mathbf{k}) =
\frac{1}{\omega-\mathbf{\sigma}\cdot\mathbf{k}+i\delta}.
\end{equation}
For simplicity, the fermion energy is approximated by
$\mathbf{\sigma}\cdot\mathbf{k} \sim |\mathbf{k}|$. To decouple the
temporal and spatial components of gauge field, it is convenient to
work in the Coulomb gauge $k_{i}a_{i}=0$. In the imaginary time
formalism, the propagator for the gauge field can now be written as
\begin{eqnarray}
D_{00}\left(\Omega_{m},\mathbf{q}\right) &=&
\frac{1}{D_{1}(\Omega_{m},\mathbf{q})},\\
D_{\ij}\left(\Omega_{m},\mathbf{q}\right) &=&
\left(\delta_{ij} -
\frac{q_{i}q_{j}}{\mathbf{q}^2}\right)\frac{1}{D_{\bot}
\left(\Omega_{m},\mathbf{q}\right)}
\end{eqnarray}
where $\Omega_{m} = 2m\pi T$ for bosonic modes with $m$ being
integers. The vacuum polarization functions
$D_{1}(\Omega_{m},\mathbf{q})$ and $D_{\bot}(\Omega_{m},\mathbf{q})$
come from the one-loop bubble diagram of Dirac fermions to the
leading order of $1/N$ expansion. In particular, the polarization
function appearing in the spatial component is given by
\begin{eqnarray}
D_{\bot}(\Omega_{m},\mathbf{q}) = D_{2}(\Omega_{m},\mathbf{q}) -
\frac{\Omega_{m}^{2}}{\mathbf{q}^{2}}D_{1}(\Omega_{m},\mathbf{q}).
\end{eqnarray}
The functions $D_{1}(\Omega_{m},\mathbf{q})$ and
$D_{2}(\Omega_{m},\mathbf{q})$ are defined as
\begin{eqnarray}
D_{1}(\Omega_{m},\mathbf{q}) &=& -NT\sum_{\omega_{n}}
\int\frac{d^{2}k}{(2\pi)^{2}}\mathrm{Tr}[G_{0}(\omega_{n},\mathbf{k})
G_{0}(\Omega_{m}+\omega_{n},\mathbf{q}+\mathbf{k})], \\
D_{2}(\Omega_{m},\mathbf{q}) &=& NT\sum_{\omega_{n}}
\int\frac{d^{2}k}{(2\pi)^{2}}\mathrm{Tr}[\sigma_{i}
G_{0}(\omega_{n},\mathbf{k})\sigma_{i}
G_{0}(\Omega_{m}+\omega_{n},\mathbf{q}+\mathbf{k})].
\end{eqnarray}
At zero temperature, $T=0$, it is straightforward to show that the
retarded polarization functions have the forms
\begin{eqnarray}
D_{1}\left(\Omega,\mathbf{q}\right) &=&
\frac{N\mathbf{q}^{2}\theta(|\mathbf{q}|-|\Omega|)}{16\sqrt{|\mathbf{q}|^{2}-\Omega^{2}}}
+ i\mathrm{sgn}\Omega \frac{N\mathbf{q}^{2}
\theta(|\Omega|-|\mathbf{q}|)}{16\sqrt{\Omega^{2}-|\mathbf{q}|^2}}, \\
D_{\bot}\left(\Omega,\mathbf{q}\right) &=&
\frac{N}{16}\theta(|\mathbf{q}|-|\Omega|)
\sqrt{|\mathbf{q}|^{2}-\Omega^{2}} -
i\mathrm{sgn}\Omega\frac{N}{16}\theta(|\Omega|-|\mathbf{q}|)
\sqrt{\Omega^{2}-|\mathbf{q}|^{2}}.
\end{eqnarray}

The fermion damping rate can be calculated by either the Fermi
golden rule or the, basically equivalent but more formal,
diagrammatic many-body technique \cite{Giuliani}. We will utilize
the latter one since it is easier to write down the self-consistent
equations using diagrammatic technique.

\subsection{Perturbative computation of fermion damping rate}

\begin{figure}[h]
\centering
    \includegraphics[width=2.4in]{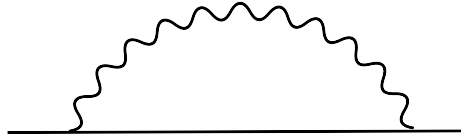}
 \caption{Fermion self-energy correction to the leading order. The solid line
 represents the massless Dirac fermion, and the wiggly
 line represents the gauge field.}
\label{fig:deltaf} %% label for entire figure
\end{figure}

We now calculate the fermion damping rate using conventional
perturbative method. To the lowest order of $1/N$ expansion, the
one-loop self-energy of Dirac fermion is given by
Fig. \ref{fig:deltaf} which can be written as
\begin{equation}
\Sigma(\omega_n,\mathbf{k})
=\Sigma_{\mathrm{L}}(\omega_{n},\mathbf{k}) +
\Sigma_{\mathrm{T}}(\omega_{n},\mathbf{k}),
\end{equation}
where
\begin{eqnarray}
\Sigma_{\mathrm{L}}(\omega_n,\mathbf{k}) &=&- T\sum_{\Omega_{m}}
\int\frac{d^2\mathbf{q}}{(2\pi)^{2}}
G_{0}(\omega_{n}+\Omega_{m},\mathbf{k+q})
D_{00}(\Omega_{m},\mathbf{q})\\
\Sigma_{\mathrm{T}}(\omega_n,\mathbf{k}) &=& T\sum_{\Omega_{m}}
\int\frac{d^2\mathbf{q}}{(2\pi)^{2}}\sigma_{i}
G_{0}(\omega_{n}+\Omega_{m},\mathbf{k+q})\sigma_{j}
D_{ij}(\Omega_{m},\mathbf{q})
\end{eqnarray}
which represent the contribution from the longitudinal and
transverse gauge field, respectively. The damping rate of massless
Dirac fermion can be obtained by making analytic continuation,
$i\omega_n\rightarrow\omega+i\delta$, as
\begin{equation}
\Sigma(\omega,\mathbf{k}) = \Sigma_{\mathrm{L}}(\omega,\mathbf{k}) +
\Sigma_{\mathrm{T}}(\omega,\mathbf{k}),
\end{equation}
and then taking the imaginary part, $\mathrm{Im}\Sigma(\omega,\mathbf{k})$.

We first consider the transverse contribution to the damping rate at
$T=0$. Using the spectral representations for Dirac fermion and
gauge boson propagators
\begin{eqnarray}
G_{0}(\omega_{n}+\Omega_{m},\mathbf{k+q}) &=&
-\int_{-\infty}^{+\infty}\frac{d\omega_1}{\pi}
\frac{\mathrm{Im}G_{0}(\omega_{1},\mathbf{k+q})}{i\omega_{n}
+ i\Omega_{m} - \omega_{1}},\\
\frac{1}{D_{\bot}(\Omega_{m},\mathbf{q})} &=&
-\int_{-\infty}^{+\infty}\frac{d\omega_2}{\pi}
\frac{\mathrm{Im}\frac{1}{D_{\bot}(\omega_{2},
\mathbf{q})}}{i\Omega_{m}-\omega_{2}},
\end{eqnarray}
the imaginary part of retarded self-energy function can be cast in the form
\begin{eqnarray}
\mathrm{Im}\Sigma_{\mathrm{T}}(\omega,\mathbf{k}) &=&
\int\frac{d^2\mathbf{q}}{(2\pi)^2}\sigma_{i}
\int_{-\infty}^{+\infty}\frac{d\omega_1}{\pi}
\mathrm{Im}[G_{0}(\omega_{1},\mathbf{k+q})]\sigma_{j}
(\delta_{ij}-\frac{q_{i}q_{j}}{\mathbf{q}^{2}})
\mathrm{Im}\left[\frac{1}{D_{\bot}(\omega_{1}-\omega,\mathbf{q})}\right]\nonumber\\
&&\times\left[n_{B}(\omega_{1}-\omega) + n_{F}(\omega_{1})\right].
\end{eqnarray}
In the $T=0$ limit, the occupation numbers simplify to
\begin{equation}
n_{B}(\omega_{1}-\omega)+n_{F}(\omega_{1}) \rightarrow
-\theta(\omega_{1})\theta(\omega-\omega_{1}).
\end{equation}
In the above expression, the imaginary retarded fermion propagator
is $\mathrm{Im}[G_{0}(\omega,\mathbf{k})] =
-\pi\delta(\omega-|\mathbf{k}|)$, while the photon propagator
has the form
\begin{equation}
\mathrm{Im}\frac{1}{D_{\bot}(\Omega,\mathbf{q})} =
\frac{16}{N\mathrm{sgn}\Omega\sqrt{\Omega^{2}-|\mathbf{q}|^{2}}}
\theta(|\Omega|-|\mathbf{q}|).
\end{equation}
After straightforward computation, the transverse fermion damping
rate is found to be
\begin{equation}
\mathrm{Im}\Sigma_{\mathrm{T}}\left(\omega,\mathbf{k}\right)
=-\frac{4}{N\pi^2}\int d^2\mathbf{q}
\frac{1}{\sqrt{\left(\omega-\left|\mathbf{k+q}\right|\right)^2-\left|\mathbf{q}\right|^2}}
\theta\left(\omega-\left|\mathbf{k+q}\right|-\left|\mathbf{q}\right|\right).
\end{equation}
To get an analytic expression, we first choose to use the frequently
used on-shell approximation $\omega \equiv \omega_{\mathbf{k}} =
\left|\mathbf{k}\right|$ and get
\begin{equation}
\mathrm{Im}\Sigma_{\mathrm{T}}\left(\omega_{\mathbf{k}}\right) \nonumber \\
= -\frac{4}{N\pi^2}\int d^2\mathbf{q}
\frac{1}{\sqrt{\left(\left|\mathbf{k}\right|-\left|\mathbf{k+q}\right|\right)^2-\left|\mathbf{q}\right|^2}}
\theta\left(\left|\mathbf{k}\right|-\left|\mathbf{k+q}\right|-\left|\mathbf{q}\right|\right).
\end{equation}
Since the step function always satisfies
$\theta\left(|\mathbf{k}|-\left|\mathbf{k+q}\right|-|\mathbf{q}|\right)\equiv
0$, we know that
$\mathrm{Im}\Sigma_{\mathrm{T}}\left(\omega_{\mathbf{k}}\right)\equiv
0$. More generally, $\theta\left(\omega-\left|\mathbf{k+q}\right| -
|\mathbf{q}|\right)\equiv 0$ for $\left|\omega\right|\leq
\left|\mathbf{k}\right|$, so that
\begin{equation}
\mathrm{Im}\Sigma_{\mathrm{T}}\left(\omega\leq\left|\mathbf{k}\right|,{\mathbf{k}}\right)\equiv
0.
\end{equation}
In order to get a finite analytic expression, we use the
zero-momentum limit $|\mathbf{k}| = 0$ and finally have
\begin{equation}
\mathrm{Im}\Sigma_{\mathrm{T}}\left(\omega\right)
=-\frac{8}{N\pi}\int_0^{\frac{\omega}{2}}d\left|\mathbf{q}\right|
\frac{\left|\mathbf{q}\right|}{\sqrt{\omega^2-2\omega\left|\mathbf{q}\right|}}
= -\frac{8\omega}{3N\pi}.
\end{equation}
This linear-in-energy expression clearly signals a marginal Fermi
liquid behavior \cite{Varma}.

This result is obtained for $\omega > 0$. When $\omega < 0$, the
expression should be $\mathrm{Im}\Sigma_{\mathrm{T}}(\omega) =
\frac{8}{3N\pi}\omega$. In general, the fermion damping rate has the
form $\mathrm{Im}\Sigma_{\mathrm{T}}(\omega) =
-\frac{8}{3N\pi}|\omega|$. Without loss of generality, we consider
only positive $\omega$ in the following. The real part of retarded
fermion self-energy can be directly obtained by the Kramers-Kronig
relation, as $\mathrm{Re}\Sigma_{\mathrm{T}}(\omega) \propto
\omega\ln\omega$.

This contribution has its own physical application. In the effective
gauge theory derived by the slave-boson approach, the gauge field
also couples to non-relativistic scalar bosons which describe the
motion of charged holons \cite{Leereview, Kim97, Kim99, Rantner}.
The scalar bosons are incompressible and thus effectively screen the
temporal component of the gauge field. So the temporal component
$a_{0}$ can be omitted, but the transverse components $\mathbf{a}$
remain unscreened and should be carefully treated. Within this
effective field theory, several interesting results have been
obtained, including the singular corrections to specific heat and
susceptibility \cite{Kim97}, and the algebraic correlation
\cite{Rantner}. The above results show that the transverse gauge
field leads to marginal Fermi liquid behavior at $T=0$.

\begin{figure}[h]
    \subfigure[]{
    \label{fig:TransversePNF1}
    \includegraphics[width=3.0in]{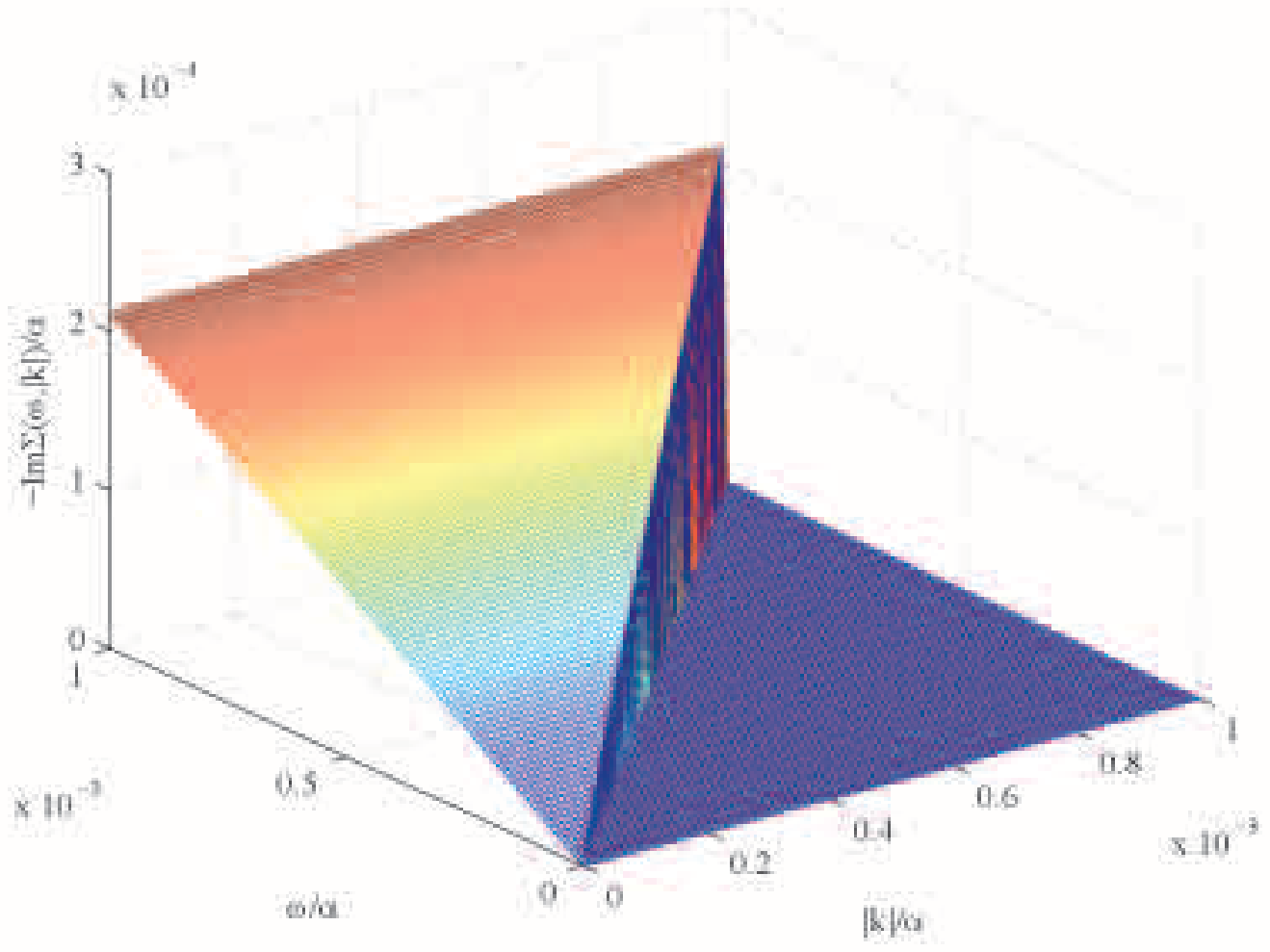}}
    \subfigure[]{
    \label{fig:TransversePNF2}
    \includegraphics[width=3.0in]{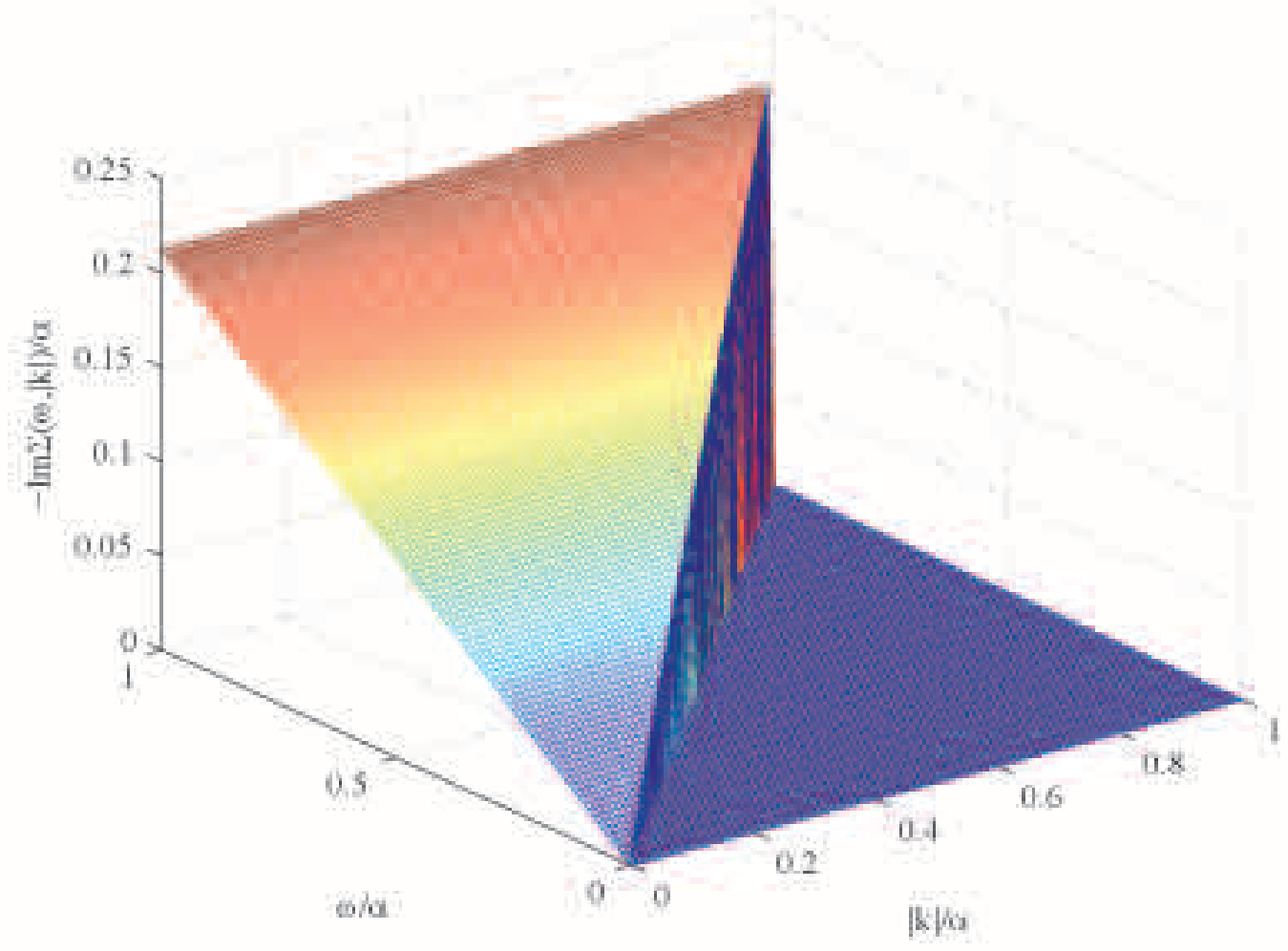}}
 \caption{Transverse contribution to damping rate of Dirac
 fermion without Maxwell term of gauge field at $T=0$.}
     \label{fig:TranversePNF}
\end{figure}

It seems impossible to get an analytical expression for the general
damping rate
$\mathrm{Im}\Sigma_{\mathrm{T}}\left(\omega,\mathbf{k}\right)$, so
we compute it by numerical skills with the results being shown in
Fig. \ref{fig:TranversePNF}.

\begin{figure}[h]
\centering
\includegraphics[width=2.4in]{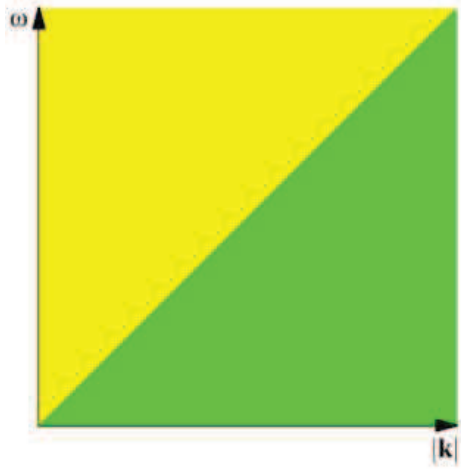}
\caption{In the yellow region where $\omega
> |\mathbf{k}|$, $\mathrm{Im}\Sigma_{\mathrm{L}}(\omega,\mathbf{k})$ is
infrared divergent. In the green region where $\omega \leq
|\mathbf{k}|$, $\mathrm{Im}\Sigma_{\mathrm{L}}(\omega,\mathbf{k}) =
0$.} \label{fig:GeneraL}
\end{figure}

The longitudinal contribution to fermion self-energy function will
be calculated analogously. The longitudinal damping rate can be
written as
\begin{equation}
\mathrm{Im}\Sigma_{\mathrm{L}}\left(\omega,\mathbf{k}\right)
=\int\frac{d^2\mathbf{q}}{\left(2\pi\right)^2}\int_0^{\omega}
\frac{d\omega_1}{\pi}\mathrm{Im}\left[G_0\left(\omega_1,\mathbf{k+q}\right)\right]
\mathrm{Im}\left[\frac{1}{D_1\left(\omega_1-\omega,\mathbf{q}\right)}\right].
\end{equation}
Using the expression for temporal gauge propagator
\begin{equation}
\mathrm{Im}\left[\frac{1}{D_{1}(\Omega,\mathbf{q})}\right] =
\frac{-16\sqrt{\Omega^{2}-|\mathbf{q}|^{2}}}{\mathrm{sgn}\Omega N
|\mathbf{q}|^{2}}\theta(|\Omega|-|\mathbf{q}|),
\end{equation}
we obtain
\begin{equation}
\mathrm{Im}\Sigma_{\mathrm{L}}\left(\omega,\mathbf{k}\right)
=-\frac{4}{N\pi^2}\int d^2\mathbf{q}
\frac{\sqrt{\left(\omega-\left|\mathbf{k+q}\right|\right)^2-\left|\mathbf{q}\right|^2}}
{\left|\mathbf{q}\right|^2}
\theta\left(\omega-\left|\mathbf{k+q}\right|-\left|\mathbf{q}\right|\right).
\end{equation}
Analogous to the transverse contribution, we can easily get
\begin{equation}
\mathrm{Im}\Sigma_{\mathrm{L}}\left(\omega\leq
|\mathbf{k}|,{\mathbf{k}}\right)\equiv 0.
\end{equation}
In the limit $|\mathbf{k}|=0$, the longitudinal damping rate is
\begin{equation}
\mathrm{Im}\Sigma_{\mathrm{L}}\left(\omega\right)
=-\frac{8}{N\pi}\int_0^{\frac{\omega}{2}}d\left|\mathbf{q}\right|
\frac{\sqrt{\omega^2-2\omega\left|\mathbf{q}\right|}}
{\left|\mathbf{q}\right|}. \nonumber
\end{equation}
Obviously there appears a serious infrared divergence. In fact,
$\mathrm{Im}\Sigma_{\mathrm{L}}\left(\omega,\mathbf{k}\right)$ is
always infrared divergent for $\omega>\left|\mathbf{k}\right|$. The
general, $\mathrm{Im}\Sigma_{\mathrm{L}}(\omega,\mathbf{k})$ can be
written as
\begin{equation}
\mathrm{Im}\Sigma_{\mathrm{L}}\left(\omega,\mathbf{k}\right)
=-\frac{4}{N\pi^2}\int_0^{+\infty}d\left|\mathbf{q}\right|
F_{1}\left(|\mathbf{q}|\right)
\end{equation}
with
\begin{equation}
F_{1}\left(|\mathbf{q}|\right) =\int_0^{2\pi}d\varphi
\frac{\sqrt{\left(\omega-\left|\mathbf{k+q}\right|\right)^2-|\mathbf{q}|^2}}
{|\mathbf{q}|}
\theta\left(\omega-\left|\mathbf{k+q}\right|-|\mathbf{q}|\right).
\end{equation}
Where $\varphi$ is the angle between $\mathbf{q}$ and $\mathbf{k}$.
When $\omega > |\mathbf{k}|$,
\begin{equation}
\lim_{|\mathbf{q}|\rightarrow 0}F_{1}\left(|\mathbf{q}|\right) =
\frac{2\pi\left(\omega-\left|\mathbf{k}\right|\right)}
{\left|\mathbf{q}\right|},
\end{equation}
which is divergent in the infrared region. The
general damping rate
$\mathrm{Im}\Sigma_{\mathrm{L}}(\omega,\mathbf{k})$ are shown in
Fig. \ref{fig:GeneraL}.

Here, an infrared divergence appears in the expression of fermion
damping rate due to the singular gauge interaction. Such divergent
damping rate is surely not well-defined at $T=0$. This does not
imply that the damping rate itself diverges, but reflects the
inefficiency of naive perturbation computation.

We now extend the above consideration to finite temperature, $T \gg
\omega$. We first calculate the longitudinal damping rate of Dirac
fermions. After a series of manipulations, in the limit
$\left|\mathbf{k}\right|=0$ ,the longitudinal component of imaginary
self-energy function is written in the form
\begin{eqnarray}
\mathrm{Im}\Sigma_{\mathrm{L}}(\omega,T) &=&
\frac{1}{2\pi}\int_{0}^{+\infty} d|\mathbf{q}||\mathbf{q}|
\mathrm{Im}\left[\frac{1}{D_{1}(|\mathbf{q}|-\omega,\mathbf{q},T)}\right]
\left[n_{B}(|\mathbf{q}|-\omega) + n_{F}(|\mathbf{q}|)\right].
\end{eqnarray}
Notice that the occupation numbers $n_{B} \left(|\mathbf{q}| - \omega\right)$ and
$n_{F}\left(|\mathbf{q}|\right)$ damp exponentially with
$\frac{|\mathbf{q}|}{T}$, so the dominant contribution of the
integral comes from the domain $|\mathbf{q}| < T$. Hence the
ultraviolet cutoff of the integral can be set to be $T$. In the high
temperature limit, it is convenient to make the simplifications
\begin{eqnarray}
n_{B}(|\mathbf{q}|-\omega) &\approx&
\frac{T}{|\mathbf{q}|-\omega}, \nonumber \\
n_{F}(|\mathbf{q}|) &\approx& \frac{1}{2}.
\end{eqnarray}
The polarization function
$D_{1}(|\mathbf{q}|-\omega,|\mathbf{q}|,T)$ in the limit $T \gg
\omega$ has a very complicated expression, which makes analytical
computation difficult. To keep the analytic tractability, it is
necessary to make some approximations. To simplify the analysis, we
divide the whole domain of $|\mathbf{q}|$ into three sections and
then perform the momentum integration respectively. Specifically, we
decompose the longitudinal damping rate into the following three
parts:
\begin{eqnarray}
\mathrm{Im}\Sigma_{\mathrm{L}}(\omega,T) &=&
\frac{1}{2\pi}\left(\int_{0}^{\frac{\omega}{2}}+\int_{\frac{\omega}{2}}^{q^{*}}
+\int_{q*}^{T}\right) d|\mathbf{q}||\mathbf{q}|
\mathrm{Im}\left[\frac{1}{D_{1}(|\mathbf{q}|-\omega,\mathbf{q},T)}\right]
\left[n_{B}(|\mathbf{q}|-\omega) +
n_{F}(|\mathbf{q}|)\right] \nonumber \\
&=& I_{1}+I_{2}+I_{3}.
\end{eqnarray}
Here, the variable $q^{*}$ is a particularly chosen quantity between
$\omega$ and $T$ which we employ to simply the calculation. The key
motivation to introduce this quantity is to specify the most
important contribution of the integration. In principle, we can
employ any $q^{*}$ between $\omega$ and $T$. In this paper, we
assume that $q^{*}$ satisfy
\begin{equation}
\sqrt{\frac{T}{q^*}}\gg\frac{q^*}{\omega}\gg 1.
\end{equation}
The polarization functions
are given by the following expressions.
When $0<\left|\mathbf{q}\right|<\frac{\omega}{2}$,
\begin{eqnarray}
&&\mathrm{Re}D_{1}\left(\left|\mathbf{q}\right|-\omega,\left|\mathbf{q}\right|,T\right)
\approx C_{1}T-C_{2}
\frac{T\left(\omega-\left|\mathbf{q}\right|\right)}
{\sqrt{\omega^2-2\omega\left|\mathbf{q}\right|}}
\\
&&\mathrm{Im}D_{1}\left(\left|\mathbf{q}\right|-\omega,\left|\mathbf{q}\right|,T\right)
\approx-C_{3}\frac{\left|\mathbf{q}\right|^2\left(\omega-\left|\mathbf{q}\right|\right)}
{T\sqrt{\omega^2-2\omega\left|\mathbf{q}\right|}}
\end{eqnarray}
When
$\frac{\omega}{2}<\left|\mathbf{q}\right|<q^{*}$,
\begin{eqnarray}
&&\mathrm{Re}D_{1}\left(\left|\mathbf{q}\right|-\omega,\left|\mathbf{q}\right|,T\right)
\approx C_{1}T
+C_{4}\frac{\left|\mathbf{q}\right|^3}{T\sqrt{2\omega\left|\mathbf{q}\right|
-\omega^2}}
\\
&&\mathrm{Im}D_{1}\left(\left|\mathbf{q}\right|-\omega,\left|\mathbf{q}\right|,T\right)
\approx C_{2}\frac{T\left(\left|\mathbf{q}\right|-\omega\right)}
{\sqrt{2\omega\left|\mathbf{q}\right|-\omega^2}}
\end{eqnarray}
When $q^{*}<\left|\mathbf{q}\right|<T$,
\begin{eqnarray}
&&\mathrm{Re}D_{1}\left(\left|\mathbf{q}\right|-\omega,\left|\mathbf{q}\right|,T\right)
\approx C_{1}T
+C_{3}\frac{\left|\mathbf{q}\right|^\frac{5}{2}}{T\sqrt{2\omega}}
\\
&&\mathrm{Im}D_{1}\left(\left|\mathbf{q}\right|-\omega,\left|\mathbf{q}\right|,T\right)
\approx C_{2}\frac{T\sqrt{\left|\mathbf{q}\right|}}{\sqrt{2\omega}}
\end{eqnarray}
where $C_1=N\frac{\ln2}{\pi}$,
 $C_2=\frac{N}{\pi}\left(\frac{1}{8}+2e^{-1}\right)$,
 $C_3=\frac{N}{64}$, $C_4=\frac{N}{48\pi}$.
After direct calculations, we find
\begin{eqnarray}
&&I_{1} \propto \frac{\omega^3}{T^2}, \\
&&I_{2} \propto \frac{{q^*}^{\frac{3}{2}}}{\sqrt{\omega}}, \\
&&I_{3} \propto \sqrt{\omega T}.
\end{eqnarray}
It is easy to check that $|I_{3}| \gg |I_{2}| \gg |I_{1}|$.
Therefore, the temporal damping rate at $T \gg \omega$ has the form
\begin{equation}
\mathrm{Im}\Sigma_{\mathrm{L}}\left(\omega,T\right) \propto
\sqrt{\omega T}.
\end{equation}
This expression is surely not of the normal Fermi liquid type. It
is, however, not the standard marginal Fermi liquid behavior.

We next consider the transverse component of fermion damping rate.
Similar to treatments presented above, the momentum $|\mathbf{q}|$
should also be divided into three sections, so that in the limit $\left|\mathbf{k}\right|=0$
,the damping rate
$\mathrm{Im}\Sigma_{\mathrm{T}}(\omega,T)$ has the form
\begin{eqnarray}
\mathrm{Im}\Sigma_{\mathrm{T}}(\omega,T) &=&-
\frac{1}{2\pi}\left(\int_{0}^{\frac{\omega}{2}}+\int_{\frac{\omega}{2}}^{q^{*}}
+\int_{q*}^{T}\right)d|\mathbf{q}| |\mathbf{q}|
\mathrm{Im}\left[\frac{1}{D_{\bot}(|\mathbf{q}|-\omega,\mathbf{q},T)}\right]
\left[n_{B}(|\mathbf{q}|-\omega) +
n_{F}(|\mathbf{q}|)\right] \nonumber \\
&=&I'_{1}+I'_{2}+I'_{3}.
\end{eqnarray}
The approximate expression for $D_{\bot}(|\mathbf{q}|-\omega,\mathbf{q},T)$ were
shown as follow.When $0<\left|\mathbf{q}\right|<\frac{\omega}{2}$,
\begin{eqnarray}
&&\mathrm{Re}D_{\bot}\left(\left|\mathbf{q}\right|-\omega,\left|\mathbf{q}\right|,T\right)
\approx C_{1}\frac{T\left(\omega-\left|\mathbf{q}\right|\right)^2}{\left|\mathbf{q}\right|^2}
-C_{2}\frac{T\left(\omega-\left|\mathbf{q}\right|\right)
\sqrt{\omega^2-2\omega\left|\mathbf{q}\right|}}{\left|\mathbf{q}\right|^2}
\\
&&\mathrm{Im}D_{\bot}\left(\left|\mathbf{q}\right|-\omega,\left|\mathbf{q}\right|,T\right)
\approx C_{3}\frac{\left(\omega-\left|\mathbf{q}\right|\right)
\sqrt{\omega^2-2\omega\left|\mathbf{q}\right|}}{T}
\end{eqnarray}
When $\frac{\omega}{2}<\left|\mathbf{q}\right|<q^{*}$,
\begin{eqnarray}
&&\mathrm{Re}D_{\bot}\left(\left|\mathbf{q}\right|-\omega,\left|\mathbf{q}\right|,T\right)
\approx C_{1}\frac{T\left(\left|\mathbf{q}\right|-\omega\right)^2}{\left|\mathbf{q}\right|^2}
+C'_{4}\frac{\left|\mathbf{q}\right|\sqrt{2\omega\left|\mathbf{q}\right|-\omega^2}}{T}
\\
&&\mathrm{Im}D_{\bot}\left(\left|\mathbf{q}\right|-\omega,\left|\mathbf{q}\right|,T\right)
\approx-C_{2}\frac{T\left(\left|\mathbf{q}\right|-\omega\right)
\sqrt{2\omega\left|\mathbf{q}\right|-\omega^2}}{\left|\mathbf{q}\right|^2}
\end{eqnarray}
When $q^{*}<\left|\mathbf{q}\right|<T$,
\begin{eqnarray}
&&\mathrm{Re}D_{\bot}\left(\left|\mathbf{q}\right|-\omega,\left|\mathbf{q}\right|,T\right)
\approx C_{1}T
\\
&&\mathrm{Im}D_{\bot}\left(\left|\mathbf{q}\right|-\omega,\left|\mathbf{q}\right|,T\right)
\approx-C_{2}\frac{T\sqrt{2\omega}}{\sqrt{\left|\mathbf{q}\right|}}
\end{eqnarray}
where$C'_{4}=\frac{N}{24\pi}$. After direct calculation, we can get
\begin{eqnarray}
&&I'_1\propto\frac{\omega^3}{T^2}
\\
&&I'_2\propto\frac{T^2}{\omega}
\\
&&I'_3\propto\sqrt{\omega T}
\end{eqnarray}
It is clear that $\left|I'_2\right|>>
\left|I'_3\right|>>\left|I'_1\right|$,
then we can conclude that, in the limit $T>>\omega$
\begin{equation}
\mathrm{Im}\Sigma_{\mathrm{T}}\left(\omega,T\right) \propto
\frac{T^2}{\omega}.
\end{equation}
This expression is also divergent at zero energy $\omega \rightarrow
0$.

The above calculations were carried out in the zero-momentum limit
$|\mathbf{k}|=0$. Using the on-shell approximation $\omega \equiv
\omega_{\mathbf{k}}=\left|\mathbf{k}\right|$ at finite temperature
$T\gg\omega_{\mathbf{k}}$, we find that
$\mathrm{Im}\Sigma_{\mathrm{L}}\left(\omega_{\mathbf{k}},T\right)
\propto\sqrt{\omega_{\mathbf{k}}T}$, but that
$\mathrm{Im}\Sigma_{\mathrm{T}}\left(\omega_{\mathbf{k}},T\right)$
is divergent in the infrared region. If we instead work in the
zero-energy limit $\omega = 0$ at $T\gg\left|\mathbf{k}\right|$,
then we find that
$\mathrm{Im}\Sigma_{\mathrm{L}}\left(|\mathbf{k}|,T\right) \propto
\sqrt{|\mathbf{k}|T}$ and
$\mathrm{Im}\Sigma_{\mathrm{T}}\left(|\mathbf{k}|,T\right) \propto
\sqrt{|\mathbf{k}|T}$.

\subsection{Self-consistent computation of fermion damping rate}

The perturbative results for Dirac fermion damping rate always
contain divergence, at both zero and finite temperature. It seems
difficult to eliminate these divergences by including higher order
corrections because they arise essentially from the singular gauge
interaction. It is worth pointing out that such divergence exits in
a wide range of physical problems. In the interacting electron gas,
the electron self-energy diverges at zero energy if the bare,
long-range Coulomb interaction is considered. In realistic metals,
however, the bare Coulomb potential is always replaced by a
short-ranged Yukawa potential after including the dynamical
screening effect. The Debye screening ensures the infrared safety of
the problem. In the present issue, however, the gauge interaction
remains long-ranged even after taking the dynamical screening into
account. Indeed, the gauge invariance ensures the masslessness of
the U(1) gauge boson. Analogous divergence also appears in the
perturbative computation for the damping rate of non-relativistic
spinons due to scattering by gauge field \cite{Lee92}.

There is also an infrared divergence when computing perturbatively
the self-energy of Dirac fermion in disordered potential
\cite{Lee93}. Such divergence is essentially due to the linear
spectrum of Dirac fermions. The most popular method to overcome this
divergence is to invoke the so-called self-consistent Born
approximation, which can give rise to a finite scattering rate
\cite{Lee93}. It is natural to ask whether similar self-consistent
approach can be used to eliminate the infrared divergence appearing
in the present problem. To answer this question, we replace the
internal free fermion propagator of Fig.1 by the following retarded
propagator
\begin{equation}
G(\omega,\mathbf{k}) =
\frac{1}{\omega-\mathbf{\sigma\cdot k}-i\mathrm{Im}\Sigma(\omega,T)},
\label{eqn:DressedFermion}
\end{equation}
and then construct an integral equation for the damping rate
$\mathrm{Im}\Sigma(\omega,T)$. If the perturbative expressions for
the polarization functions are used, then only trivial result can be
obtained after numerical computation. This is easy to understand by
noting the fact that the infrared divergence originates from the
singular gauge interaction.

All the above computations are based on the polarization functions
obtained from the free propagator of Dirac fermion. The feedback of
fermion damping due to gauge field is completely neglected when
calculating the dynamically screened gauge field and fermion
self-energy function. However, if the fermion damping is really
significant, as it should be in a non-Fermi liquid, its effect can
not be simply neglected. It is conceivable to speculate that the
divergences appearing in the Dirac fermion self-energy might be
eliminated by incorporating the feedback effect of fermion damping.
We now turn to such kind of self-consistent treatment of Dirac
fermions and gauge bosons. Formally, the fermion damping rate
satisfies the integral equation
\begin{eqnarray}
\mathrm{Im}\Sigma(\omega,T) &=&
\int\frac{d^2\mathbf{q}}{(2\pi)^{2}}\int_{-\infty}^{+\infty}
\frac{d\omega_{1}}{\pi}\frac{\mathrm{Im}\Sigma(\omega_{1},T)}
{\left(\omega_{1}-|\mathbf{q}|\right)^{2} +
\left(\mathrm{Im}\Sigma(\omega_{1},T)\right)^{2}}
\nonumber \\
&& \times \mathrm{Im}\left[-\frac{1}{D_{1}
\left(\omega_{1}-\omega,\mathbf{q},T\right)}+\frac{1}{D_{\bot}
\left(\omega_{1}-\omega,\mathbf{q},T\right)}\right]
\left[n_{B}(\omega_{1}-\omega) + n_{F}(\omega_{1})\right].\label{eqn:SelfConsistentE1}
\end{eqnarray}
Using the fermion propagator (\ref{eqn:DressedFermion}), the polarization functions
$\mathrm{Im}D_{1}$ and $\mathrm{Im}D_{\bot}$ can be constructed as
follows
\begin{eqnarray}
&&\mathrm{Im}D_1\left(\varepsilon,|\mathbf{q}|,T\right)\nonumber\\
&=&2N\int\frac{d^2\mathbf{k}}{(2\pi)^2}
\frac{d\omega_1}{\pi}\left\{
\mathrm{Im}\left[\frac{\omega_1-i\mathrm{Im}\Sigma (\omega_1,T)}
{\left(\omega_{1}-i\mathrm{Im}\Sigma(\omega_1,T)\right)^{2} -
|\mathbf{k}|^{2}}\right]\mathrm{Im}\left[\frac{\omega_1+\varepsilon
- i\mathrm{Im}\Sigma(\omega_1 +
\varepsilon,T)}{\left(\omega_1+\varepsilon-i\mathrm{Im}
\Sigma(\omega_1+\varepsilon,T)\right)^2 - |\mathbf{k+q}|^2}\right]\right. \nonumber \\
&&\left. + \mathbf{k\cdot
(k+q)}\mathrm{Im}\left[\frac{1}{\left(\omega_1-i\mathrm{Im} \Sigma
(\omega_1,T)\right)^2-|\mathbf{k}|^{2}}\right]
\mathrm{Im}\left[\frac{1}{\left(\omega_1+\varepsilon-i\mathrm{Im}\Sigma
(\omega_1+\varepsilon,T)\right)^{2} -
|\mathbf{k+q}|^{2}}\right]\right\} \nonumber \\
&& \times \left[n_{F}(\omega_{1}) -
n_{F}(\omega_1+\varepsilon)\right]. \label{eqn:SelfConsistentD1Im}\\ \nonumber \\
&&\mathrm{Re}D_{1}(\varepsilon,|\mathbf{q}|,T)\nonumber\\
&=&2N\int\frac{d^2\mathbf{k}}{(2\pi)^{2}} \frac{d\omega_1}{\pi}
\left\{\mathrm{Im}\left[\frac{\omega_{1} -
i\mathrm{Im}\Sigma(\omega_{1},T)}
{\left(\omega_{1}-i\mathrm{Im}\Sigma(\omega_{1},T)\right)^{2}-|\mathbf{k}|^{2}}\right]
\mathrm{Re}\left[\frac{\omega_1+\varepsilon-i\mathrm{Im}\Sigma
\left(\omega_1+\varepsilon,T\right)}{\left(\omega_1+\varepsilon-i\mathrm{Im}
\Sigma \left(\omega_1+\varepsilon,T\right)\right)^2 -
\left|\mathbf{k+q}\right|^2}\right]\right. \nonumber \\
&&\left.+\mathbf{k\cdot
(k+q)}\mathrm{Im}\left[\frac{1}{\left(\omega_1-i\mathrm{Im}
\Sigma^{\mathrm{ret}}\left(\omega_1,T\right)\right)^2-\left|\mathbf{k}\right|^2}\right]
\mathrm{Re}\left[\frac{1}{\left(\omega_1+\varepsilon-i\mathrm{Im}\Sigma^{\mathrm{ret}}
\left(\omega_1+\varepsilon,T\right)\right)^2-\left|\mathbf{k+q}\right|^2}\right]\right\}\nonumber
\\
&&\times\left[n_{F}(\omega_{1}) -
n_{F}\left(\omega_{1}+\varepsilon\right)\right],\label{eqn:SelfConsistentD1Re}
\\ \nonumber \\
&&\mathrm{Im}D_{2}\left(\varepsilon,|\mathbf{q}|,T\right)\nonumber\\
&=&-4N\int\frac{d^{2}\mathbf{k}}{(2\pi)^{2}}
\frac{d\omega_1}{\pi}\mathrm{Im}\left[\frac{\omega_{1}-i\mathrm{Im}\Sigma(\omega_{1},T)}
{\left(\omega_1-i\mathrm{Im}\Sigma(\omega_1,T)\right)^{2} -
|\mathbf{k}|^2}\right]
\mathrm{Im}\left[\frac{\omega_{1}+\varepsilon-i\mathrm{Im}\Sigma
(\omega_1+\varepsilon,T)}{\left(\omega_1+\varepsilon-i\mathrm{Im}
\Sigma(\omega_1+\varepsilon,T)\right)^{2}-|\mathbf{k+q}|^{2}}\right]
\nonumber \\
&& \times \left[n_{F}(\omega_{1}) - n_{F}(\omega_1+\varepsilon)\right], \label{eqn:SelfConsistentD2Im} \\
\nonumber \\
&&\mathrm{Re}D_{2}\left(\varepsilon,|\mathbf{q}|,T\right)\nonumber\\
&=&-4N\int\frac{d^{2}\mathbf{k}}{(2\pi)^{2}}\frac{d\omega_{1}}{\pi}
\mathrm{Im}\left[\frac{\omega_{1}-i\mathrm{Im}\Sigma(\omega_{1},T)}
{\left(\omega_{1}-i\mathrm{Im}\Sigma(\omega_{1},T)\right)^{2} -
|\mathbf{k}|^{2}}\right]
\mathrm{Re}\left[\frac{\omega_1+\varepsilon-i\mathrm{Im}\Sigma
(\omega_1+\varepsilon,T)}{\left(\omega_1+\varepsilon-i\mathrm{Im}
\Sigma(\omega_1+\varepsilon,T)\right)^{2} -
|\mathbf{k+q}|^{2}}\right] \nonumber \\
&& \times \left[n_{F}(\omega_{1}) -
n_{F}(\omega_1+\varepsilon)\right]\label{eqn:SelfConsistentD2Re},
\end{eqnarray}
where
\begin{eqnarray}
D_{\bot}(\varepsilon,|\mathbf{q}|,T) =
D_{2}(\varepsilon,|\mathbf{q}|,T)+\frac{\varepsilon^2}{|\mathbf{q}|^{2}}
D_{1}(\varepsilon,|\mathbf{q}|,T).\label{eqn:SelfConsistentDBot}
\end{eqnarray}

These equations appear to be very complicated and hard to solve. To
get the fermion damping rate from these coupled integral equations,
we find it is convenient to employ a simple dimensional analysis
similar to that used by Vojta \emph{et} \emph{al.} \cite{Vojta}.
After dividing all momenta, energy, and self-energy function by
$\omega$ at $T = 0$ and by $T$ at finite $T$, it is found that
$\mathrm{Im}\Sigma(\omega) \propto \omega$ and $\mathrm{Im}\Sigma(T)
\propto T$ respectively. Intuitively, this re-scaling procedure
reflects the typical behavior of a marginal Fermi liquid. However,
if we define
\begin{eqnarray}
\frac{\mathrm{Im}\Sigma(\omega)}{\omega} &=& A, \\
\frac{\mathrm{Im}\Sigma(T)}{T} &=& B,
\end{eqnarray}
then numerical calculation find no convergent solutions for $A$ and
$B$. It turns out that, although formally the damping rate depends
linearly on $\omega$ at $T=0$ and on $T$ at $\omega=0$, a divergence
appears in the regime where the energy of polarization functions
vanishes. Specifically, this divergence emerges when $\omega_{1}
\rightarrow \omega$ in the fermion damping rate equation
(\ref{eqn:SelfConsistentE1}). Since the functions $D_{1}$ and
$D_{\bot}$ appear in equation ({\ref{eqn:SelfConsistentE1}}) as
denominators, the fermion damping rate actually diverges as the
energy of polarization functions vanishes. This qualitative analysis
is confirmed by the numerical computations.

In the above, we show that it is hard to get meaningful results of
fermion damping rate within both perturbation theory and
self-consistent treatment when the gauge field has no explicit
Maxwell term. There is always some kind of divergence. All these
divergences originate from the dynamically screened propagator of
gauge field. The fermion damping rate seems not to be a well-defined
quantity, at least under the approximations considered in the above.

\section{Fermion damping rate in QED$_{3}$ with Maxwell term\label{sec:WithMaxwellItem}}

In this section, we study the QED$_{3}$ theory with explicit Maxwell
term. According to the standard framework of relativistic quantum
field theory, it is natural to keep such kinetic term in the
Lagrangian. The QED$_{3}$ with a Maxwell term itself is very
interesting and has been studied extensively in the past twenty
years (for a review, see \cite{Appelquist04}). It also has direct
applications in condensed matter physics. In the context of
underdoped high temperature superconductors, an effective QED$_{3}$
theory was derived to model the unusual physics after carefully
considering the phase fluctuations \cite{Franz, Herbut}. There is a
Maxwell term in this kind of QED$_{3}$ theory. In the following, we
will include the Maxwell term for gauge field in the Lagrangian of
QED$_{3}$ and re-calculate the fermion damping rate.

\begin{figure}[h]
    \subfigure[]{
    \label{fig:TransversePF1}
    \includegraphics[width=3.0in]{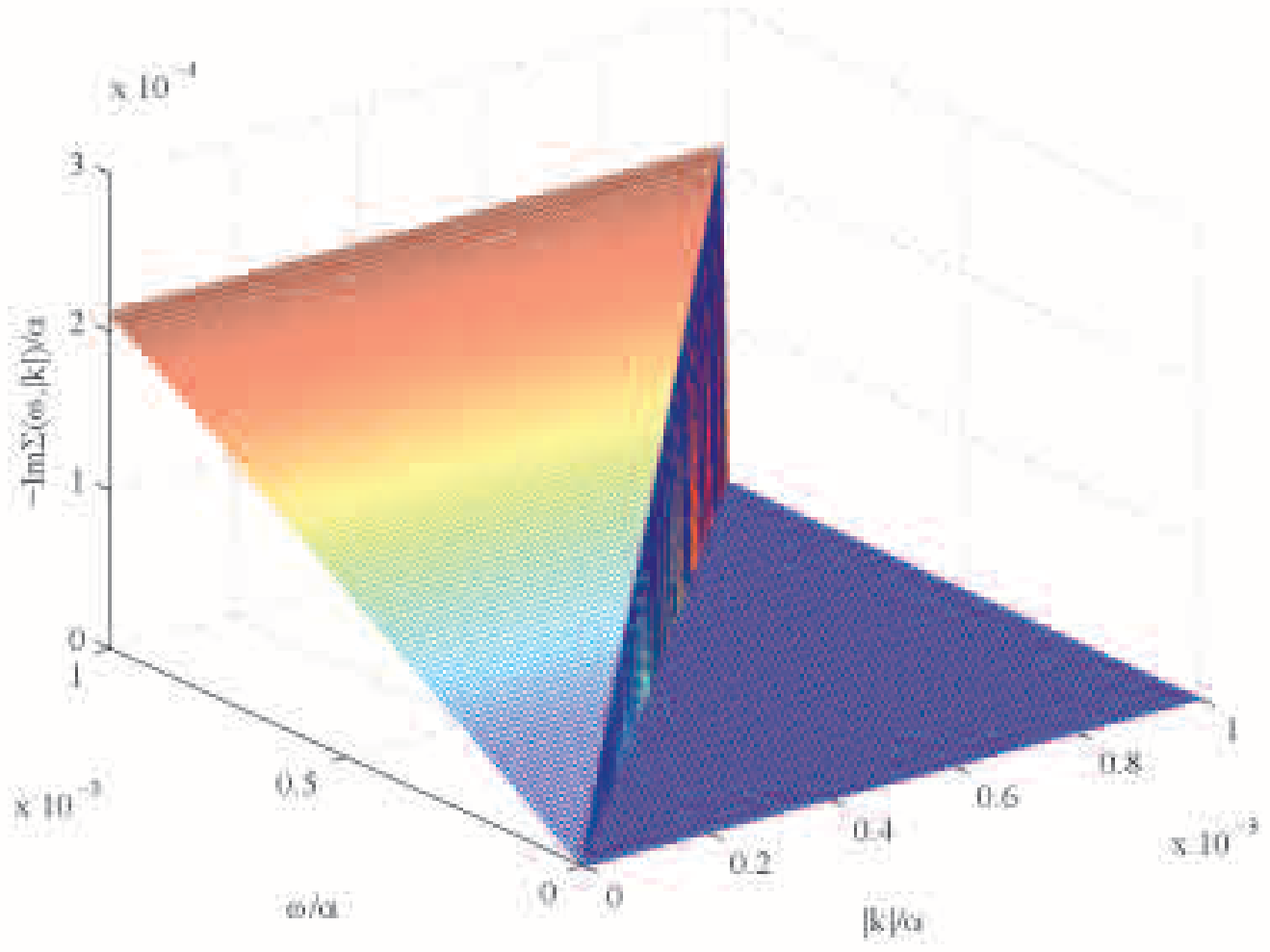}}
    \subfigure[]{
    \label{fig:TransversePF2}
    \includegraphics[width=3.0in]{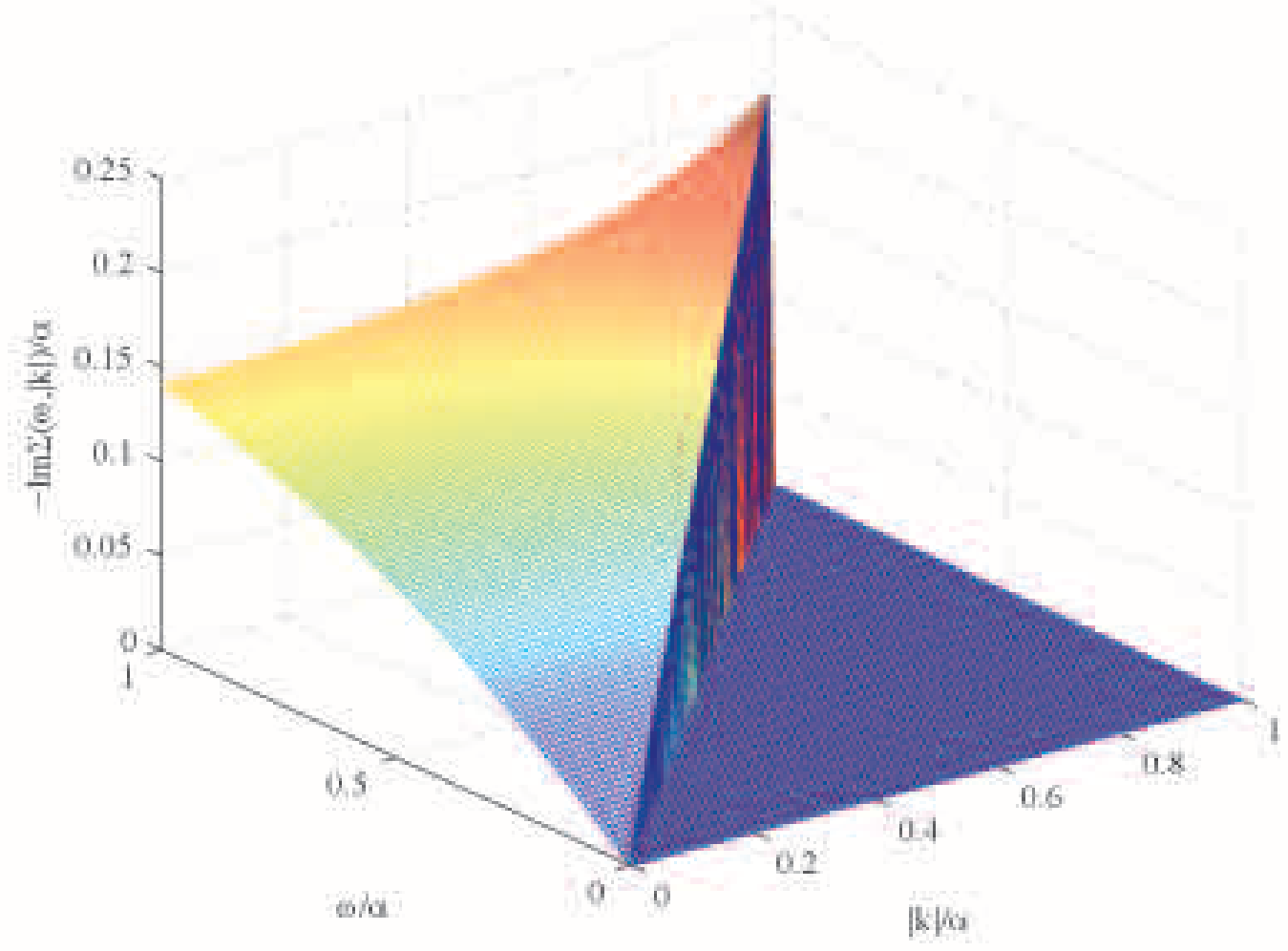}}
 \caption{Transverse contribution to damping rate for Dirac
 fermion in the presence of Maxwell term.}
     \label{fig:TranversePF}
\end{figure}

In the presence of Maxwell term, the propagators for gauge field are
\begin{eqnarray}
D_{00}(\Omega_{m},\mathbf{q}) &=&
\frac{1}{\left|\mathbf{q}\right|^{2} +
D_{1}(\Omega_m,\mathbf{q})}, \\
D_{ij}(\Omega_{m},\mathbf{q}) &=&
\left(\delta_{ij}-\frac{q_iq_j}{\mathbf{q}^2}\right)
\frac{1}{|\mathbf{q}|^{2}+\Omega_{m}^{2} +
D_\bot\left(\Omega_{m},\mathbf{q}\right)}.
\end{eqnarray}
After perturbative computations, we find the following transverse damping rate
\begin{eqnarray}
\mathrm{Im}\Sigma_{\mathrm{T}}\left(\omega,\mathbf{k}\right)
&=&-\frac{4N}{\pi^2}\int d^2\mathbf{q}
\frac{\sqrt{\left(\omega-\left|\mathbf{k+q}\right|\right)^2 -
\left|\mathbf{q}\right|^2}}{256\left(\left(\omega-|\mathbf{k+q}|\right)^2-|\mathbf{q}|^2\right)^2
+N^2\left(\left(\omega-\left|\mathbf{k+q}\right|\right)^2-\left|\mathbf{q}\right|^2\right)}
\nonumber \\
&& \times\theta\left(\omega-|\mathbf{k+q}|-|\mathbf{q}|\right).
\end{eqnarray}
Since
$\theta\left(\omega-\left|\mathbf{k+q}\right|-|\mathbf{q}|\right)\equiv
0$ when $|\omega|\leq \left|\mathbf{k}\right|$, we still have
$\mathrm{Im}\Sigma_{\mathrm{T}}\left(\omega\leq
\left|\mathbf{k}\right|,{\mathbf{k}}\right)\equiv0$. In the limit
$\left|\mathbf{k}\right|=0$, we get
\begin{eqnarray}
\mathrm{Im}\Sigma_{\mathrm{T}}(\omega) &=&
-\frac{1}{4\pi}\arctan(\frac{16\omega}{N})+ \frac{N}{64\pi\omega} -
\frac{N}{64\pi\omega}\frac{N}{16\omega} \arctan(\frac{16\omega}{N}).
\end{eqnarray}
In the low-energy regime, $\omega\rightarrow 0$, it reduces to
\begin{equation}
\mathrm{Im}\Sigma_{\mathrm{T}}(\omega) =
-\frac{8|\omega|}{3N\pi},
\end{equation}
which coincides with the standard behavior of marginal Fermi liquid.
The general
$\mathrm{Im}\Sigma_{\mathrm{T}}\left(\omega,\mathbf{k}\right)$ can
only be calculated by numerical methods, with results shown in
Fig. \ref{fig:TranversePF}.

On the other hand, the longitudinal damping rate is found to be
\begin{eqnarray}
\mathrm{Im}\Sigma_{\mathrm{L}}\left(\omega,\mathbf{k}\right)
=-\frac{4N}{\pi^2}\int d^2\mathbf{q}
\frac{\sqrt{\left(\omega-\left|\mathbf{k+q}\right|\right)^2-\left|\mathbf{q}\right|^2}}
{256\left(\left(\omega-\left|\mathbf{k+q}\right|\right)^2-\left|\mathbf{q}\right|^2\right)
\left|\mathbf{q}\right|^2+N^2\left|\mathbf{q}\right|^2}
\theta\left(\omega-\left|\mathbf{k+q}\right|-\left|\mathbf{q}\right|\right)
\end{eqnarray}
It is easy to prove that
$\mathrm{Im}\Sigma_{\mathrm{L}}\left(\omega\leq
\left|\mathbf{k}\right|,{\mathbf{k}}\right)\equiv0$. In the limit
$\left|\mathbf{k}\right|=0$,
\begin{equation}
\mathrm{Im}\Sigma_{\mathrm{L}}\left(\omega\right)
=-\frac{8N}{\pi}\int_{0}^{\frac{\omega}{2}}d\left|\mathbf{q}\right|\frac{1}{\left|\mathbf{q}\right|}
\left[\frac{\sqrt{\omega^2-2\omega\left|\mathbf{q}\right|}}
{256\left(\omega^2-2\omega\left|\mathbf{q}\right|\right)+N^2}\right]
\end{equation}
This is still divergent in the infrared region. The general
$\mathrm{Im}\Sigma_{\mathrm{L}}\left(\omega,\mathbf{k}\right)$ ban
be written as
\begin{eqnarray}
\mathrm{Im}\Sigma_{\mathrm{L}}\left(\omega,\mathbf{k}\right)
=-\frac{4N}{\pi^2}\int_0^{+\infty}d\left|\mathbf{q}\right|
F_{2}\left(\left|\mathbf{q}\right|\right)
\end{eqnarray}
where
\begin{equation}
F_{2}\left(\left|\mathbf{q}\right|\right)=
\int_0^{2\pi}d\varphi\frac{\sqrt{\left(\omega-\left|\mathbf{k+q}\right|\right)^2-\left|\mathbf{q}\right|^2}}
{256\left(\left(\omega-\left|\mathbf{k+q}\right|\right)^2-\left|\mathbf{q}\right|^2\right)
\left|\mathbf{q}\right|+N^2\left|\mathbf{q}\right|}
\theta\left(\omega-\left|\mathbf{k+q}\right|-\left|\mathbf{q}\right|\right)
\end{equation}
When $\omega > |\mathbf{q}|$, we have
\begin{equation}
\lim_{\left|\mathbf{q}\right|\rightarrow0}F_{2}\left(|\mathbf{q}|\right) \\
=
\frac{1}{|\mathbf{q}|}\left[\frac{2\pi\left(\omega-|\mathbf{k}|\right)}
{256\left(\omega-|\mathbf{k}|\right)^2+N^2}\right]. \nonumber
\end{equation}
Thus we can easily see that
$\mathrm{Im}\Sigma_{\mathrm{L}}(\omega,\mathbf{k})$ is infrared
divergent when $|\omega|>\left|\mathbf{k}\right|$.

Due to this divergence, the longitudinal fermion damping rate is
still ill-defined at zero temperature $T=0$, even in the presence of
the Maxwell term.

At finite temperature, after tedious calculations we find that the
Dirac fermion damping rate has the following features: If we
calculate in the limit $\left|\mathbf{k}\right|=0$ at $T\gg\omega$,
then $\mathrm{Im}\Sigma_{\mathrm{L}}\left(\omega,T\right) \propto
\sqrt{\omega T}$ and
$\mathrm{Im}\Sigma_{\mathrm{T}}\left(\omega,T\right)\propto
T/\omega$. If we work in the on-shell approximation
$\omega=\omega_{\mathbf{k}}=\left|\mathbf{k}\right|$ at
$T\gg\omega_{\mathbf{k}}$, then we find that
$\mathrm{Im}\Sigma_{\mathrm{L}}\left(\omega_{\mathbf{k}},T\right)
\propto\sqrt{\omega_{\mathbf{k}}T}$ and that
$\mathrm{Im}\Sigma_{\mathrm{T}}\left(\omega_{\mathbf{k}},T\right)$
is divergent in the infrared region. In the zero-energy limit
$\omega=0$ with $T \gg |\mathbf{k}|$, we have
$\mathrm{Im}\Sigma_{\mathrm{L}}\left(|\mathbf{k}|,T\right) \propto
\sqrt{\left|\mathbf{k}\right|T}$ and
$\mathrm{Im}\Sigma_{\mathrm{T}}\left(|\mathbf{k}|,T\right) \propto
\sqrt{\left|\mathbf{k}\right|T}$.

All the above calculations are obtained in the Coulomb gauge, which
separates the longitudinal and transverse components of the gauge
field completely. The same calculations can be done similarly by
choosing another gauge. After straightforward computation
\cite{Wang09}, we find that the fermion damping rate is still
divergent in a general gauge when obtained at the perturbative
level, no matter the Maxwell term of gauge field is present or not.
This implies that the existence of divergence in perturbative
expansion is a universal feature of QED$_3$, rather than just a
gauge artifact.

The marginal Fermi liquid behavior of fermion damping rate was
claimed previously by Franz and Tesanovic without providing
computational details \cite{Franz}. However, the detailed
calculations show that the fermion damping rate is an ill-defined
quantity at the perturbative level because divergence appears at
both the $T=0$ and $T\gg \omega$ limits.

\begin{figure}[h]
    \subfigure[]{
    \label{fig:ZeroT}
    \includegraphics[width=3.2in]{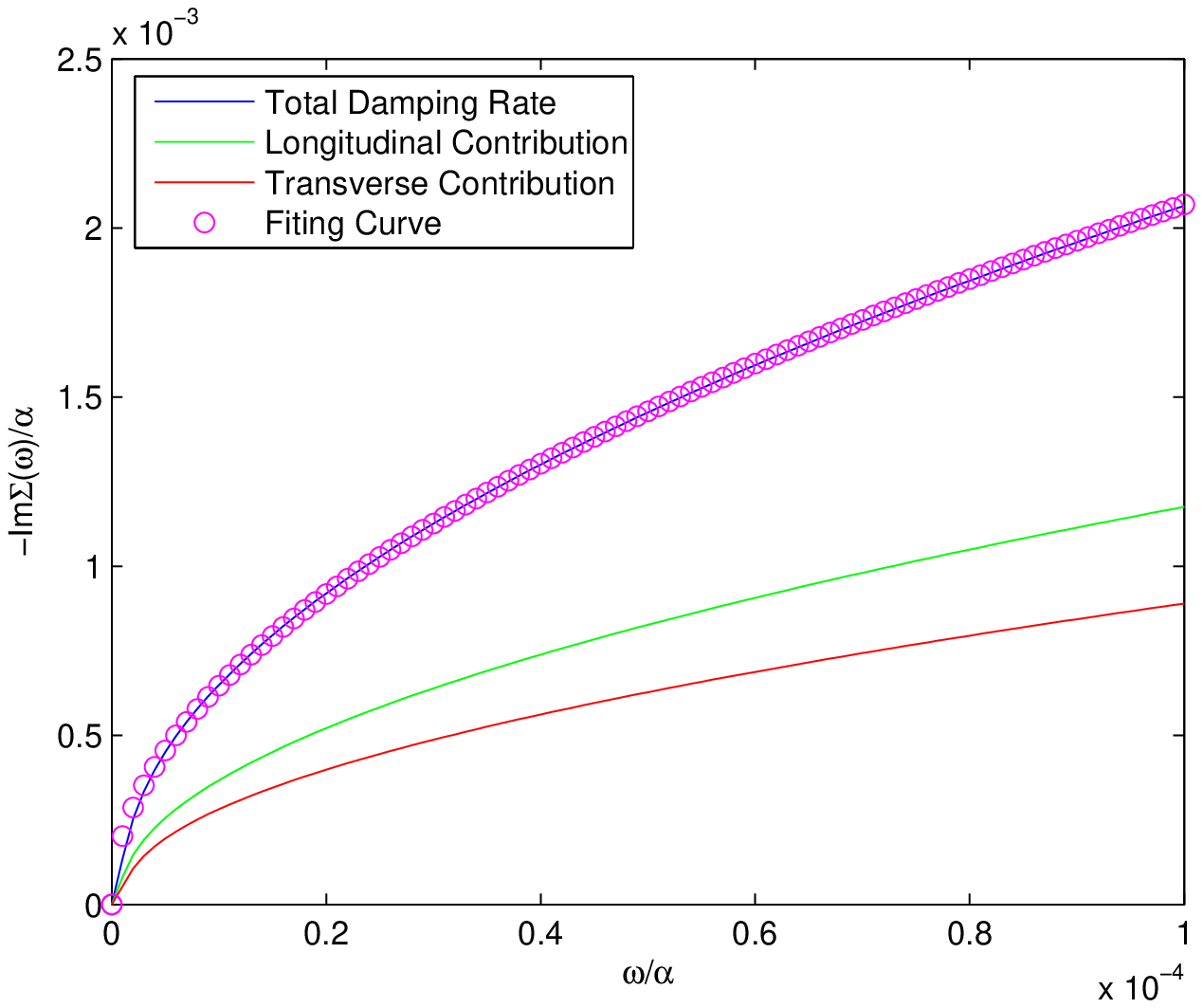}}
    \subfigure[]{
   \includegraphics[width=3.2in]{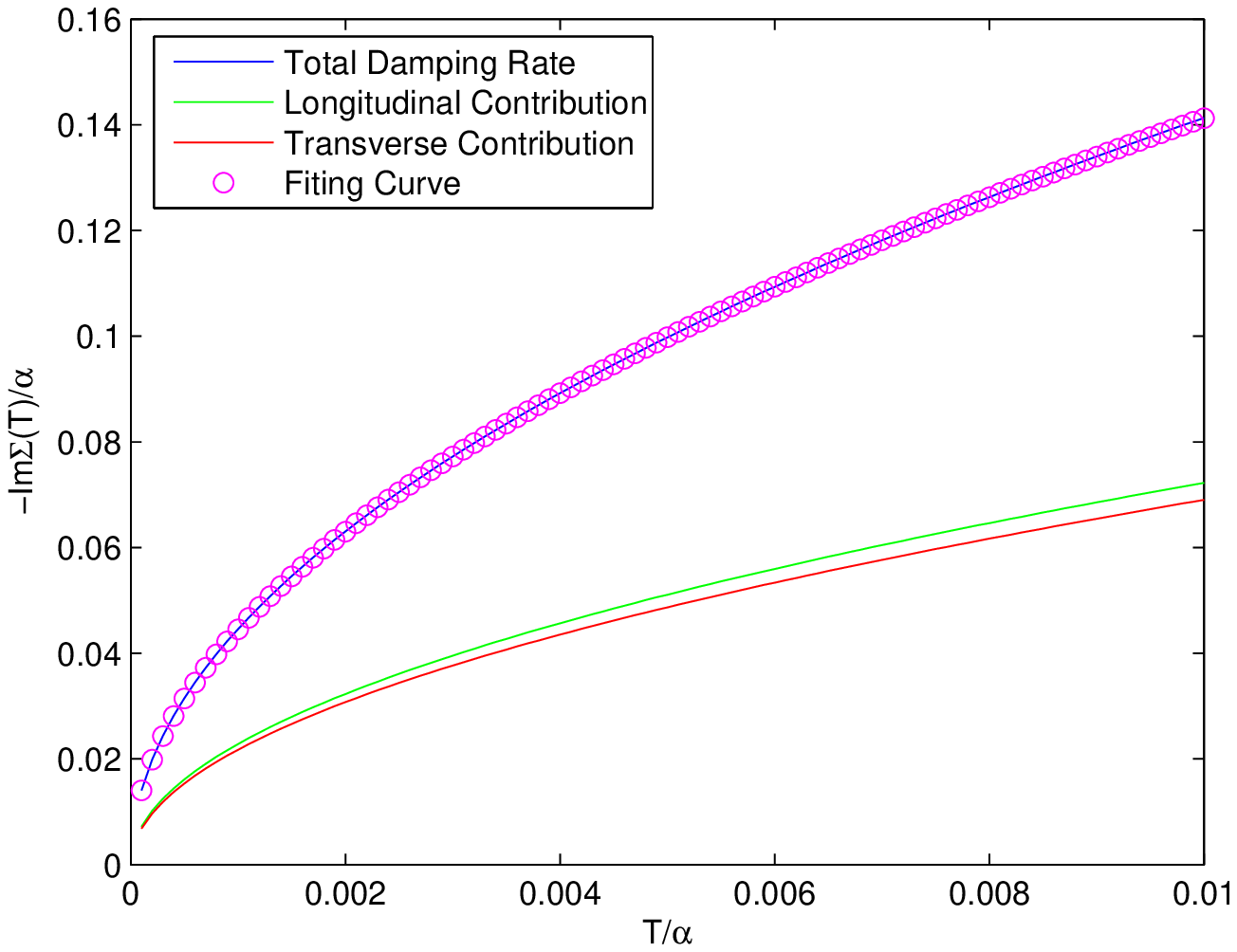}
    \label{fig:FinitT}}
\caption{(a) Damping rate of massless Dirac fermions at zero
temperature $T=0$. The fitting curve is
$\mathrm{Im}\Sigma\left(\omega\right)/\alpha =
0.2170(\omega/\alpha)^{0.5051}$. Here the energy scale $\alpha$ is
defined as $\alpha = Ne^{2}/8$ (we have let $e=1$ in this paper),
which was first introduced in Ref. \cite{Appelquist88}. (b) Damping
rate of massless Dirac fermions on Fermi level at finite
temperature. The fitting curve is
$\mathrm{Im}\Sigma\left(T\right)/\alpha =
1.4220(T/\alpha)^{0.5014}$.}
%% label for entire figure
\end{figure}

We next turn to the self-consistent treatment of fermion damping
rate in the presence of Maxwell term. To compare with the results
presented above, we also choose to work in the Coulomb gauge. As in
the last section, we include both the real and imaginary parts of
the vacuum polarizations when writing the integral equation for the
fermion damping rate
\begin{eqnarray}
&&\mathrm{Im}\Sigma(\omega,T)\nonumber\\
&=&\frac{1}{2\pi^{2}}\int_{0}^{+\infty} d|\mathbf{q}|
\int_{-\infty}^{+\infty}d\omega_{1}
\frac{\mathrm{Im}\Sigma(\omega_1,T)}{\left(\omega_{1}-|\mathbf{q}|\right)^{2}
+\left(\mathrm{Im}\Sigma(\omega_{1},T)\right)^{2}} \nonumber \\
&& \times \left\{-\mathrm{Im}\left[\frac{1}{|\mathbf{q}|^{2}+
D_{1}\left(\omega_{1}-\omega,|\mathbf{q}|,T\right)}\right]
+\mathrm{Im}\left[\frac{1}{
\left|\mathbf{q}\right|^{2}-\left(\omega_1-\omega\right)^2+D_{\bot}^{\mathrm{ret}}
\left(\omega_1-\omega,\left|\mathbf{q}\right|,T\right)}\right]\right\}
\nonumber \\
&& \times \left[n_{B}(\omega_{1}-\omega) + n_{F}(\omega_{1})\right],
\end{eqnarray}
where the functions $D_{1}$ and $D_{\bot}$ are given by equations
(\ref{eqn:SelfConsistentD1Im}-\ref{eqn:SelfConsistentDBot}) in the
last section. The kinetic term of gauge field eliminated the
divergence brought by the polarization functions. At the same time,
the energy $\omega$ can no longer be scaled out due to the kinetic
gauge term. Thus the fermion damping rate is not expected to display
the marginal Fermi liquid behavior. After numerically solving these
coupled integral equations, we find that the total fermion damping
rate is
\begin{equation}
\mathrm{Im}\Sigma(\omega) \propto \sqrt{\omega}
\end{equation}
at zero temperature $T=0$. The numerical results at $T=0$ are
presented in Fig. \ref{fig:ZeroT}. The computation of fermion
damping rate at finite temperature follows the same procedure as
presented above. The numerical computations find the following
fermion damping rate
\begin{equation}
\mathrm{Im}\Sigma(T) \propto \sqrt{T}
\end{equation}
in the limit $T \gg \omega$. The numerical results for this limit
are presented in Fig. \ref{fig:FinitT}. As a summary, the damping
rate of massless Dirac fermions is
\begin{equation}
\mathrm{Im}\Sigma(\omega,T) \propto
\mathrm{max}(\sqrt{\omega},\sqrt{T})
\end{equation}
due to scattering by the U(1) gauge field in two spatial dimensions.
This is certainly a non-Fermi liquid like behavior since the fermion
damping rate would be $\mathrm{Im}\Sigma(\omega,T) \propto
\mathrm{max}(\omega^{2},T^{2})$ (or with higher powers) in a normal
Fermi liquid.

This treatment is essentially the analogy of Eliashberg theory of
superconductors with strong electron-phonon interaction. The
propagators of fermions and intermediate bosons are
self-consistently coupled while the vertex corrections are simply
ignored. In the standard Eliashberg theory of electron-phonon
system, the vertex corrections can be safely neglected since they
are suppressed by the small parameter $m/M$, where $m$ and $M$ are
the electron mass and nuclei mass respectively. This is nothing but
the Migdal theorem \cite{Abrikosov}. In the present problem, there
is no similar mass scales since both Dirac fermions and gauge bosons
are massless. However, we do have a small expansion parameter $1/N$.
The vertex corrections are suppressed by the factor $1/N$. In the
large $N$ limit, our ignorance of the vertex corrections is valid
(Polchinksi used the same argument when studying the fermion
self-energy due to gauge interaction, see his paper in Ref.
\cite{Nayak}). The validity of our results for small $N$, such as
$N=2$, is not clear currently.

Unfortunately, at present we are unable to get a well-defined result
for the momentum dependence of fermion damping rate from the same
self-consistent treatment. The reason is that the coupled equations
become very complicated after including the momentum dependence. The
numerical results are much less reliable as those at zero momentum.

\section{Marginal Fermi liquid behavior and Coulomb interaction\label{Sec:CoulombInteraction}}

The marginal Fermi liquid was proposed by Varma \emph{et} \emph{al.}
at 1989 to understand some of the highly unusual experimental facts
in the normal state of high temperature superconductors on
phenomenological grounds \cite{Varma}. Since then, a lot of efforts
have been devoted to deriving this phenomenological theory from
certain microscopic models \cite{Varma2002, Lee92, Nayak, Franz}.
Gauge theory has long been considered as one possible candidate
\cite{Lee92, Nayak, Franz}. Within the ordinary perturbative theory,
the transverse damping rate of Dirac fermions has a linear
dependence on energy at $T=0$, which is the standard behavior of
marginal Fermi liquid. However, this result can not be trusted
because of the appearance of divergence in the longitudinal
component at $T=0$ and in the transverse component at finite
temperature. More careful theoretical and numerical computations
show that the Dirac fermions exhibit non-Fermi, rather than marginal
Fermi, liquid behavior.

However, it is possible to find signature of marginal Fermi liquid
within some models those are in form analogous to the U(1) gauge
field theory. For instance, we now consider the long-range Coulomb
interaction between Dirac fermions, which remains unscreened because
of the vanishing density of states at the Fermi level. In some
sense, the Coulomb potential may be regarded as the temporal
component of a U(1) gauge field. There is, however, a subtle
difference, which leads to important consequence. The dynamically
screened Coulomb interaction can be formally written as
\begin{eqnarray}
D_{\mathrm{C}}(i\Omega_{m},\mathbf{q}) &=& \frac{1}{\frac{|\mathbf{q}|}{\lambda}
+ D_{1}(i\Omega_{m},\mathbf{q})},
\end{eqnarray}
where the $|\mathbf{q}|$ term comes from the bare, instantaneous
Coulomb potential and the dimensionless parameter is defined as
$\lambda = \frac{2\pi}{\epsilon_{0}}$ with $\epsilon_{0}$ being the
dielectric constant. At zero temperature, the damping rate of Dirac
fermions has the form
\begin{equation}
\mathrm{Im}\Sigma_{\mathrm{C}}\left(\omega,\mathbf{k}\right)
=-\frac{4N}{\pi^2}\int d^2\mathbf{q}
\frac{\sqrt{\left(\omega-\left|\mathbf{k+q}\right|\right)^2-\left|\mathbf{q}\right|^2}}
{256\left(\left(\omega-\left|\mathbf{k+q}\right|\right)^2-\left|\mathbf{q}\right|^2\right)
\frac{1}{\lambda^2}+N^2\left|\mathbf{q}\right|^2}
\theta\left(\omega-\left|\mathbf{k+q}\right|-\left|\mathbf{q}\right|\right)
\end{equation}
Here, we also have $\mathrm{Im}\Sigma_{\mathrm{C}}\left(\omega\leq
\left|\mathbf{k}\right|,{\mathbf{k}}\right)\equiv 0$. In the limit
$\left|\mathbf{k}\right|=0$, we get
\begin{equation}
\mathrm{Im}\Sigma_{\mathrm{C}}\left(\omega\right) =
-\frac{8}{N\pi}C\left(\lambda\right)\omega
\end{equation}
with
\begin{equation}
C\left(\lambda\right)=\int_{0}^{1}dx \frac{x\sqrt{1-x}}
{\left(\frac{32}{\lambda N}\right)^2\left(1-x\right)+x^2}.
\end{equation}
This damping rate is free of divergence and can be identified as the
behavior of a marginal Fermi liquid. It agrees with the results
obtained in the context of graphene \cite{Gonzalez, DasSarma,
Khveshchenko011}.

\begin{figure}[h]
    \subfigure[]{
    \label{fig:CoulomP1}
    \includegraphics[width=3.0in]{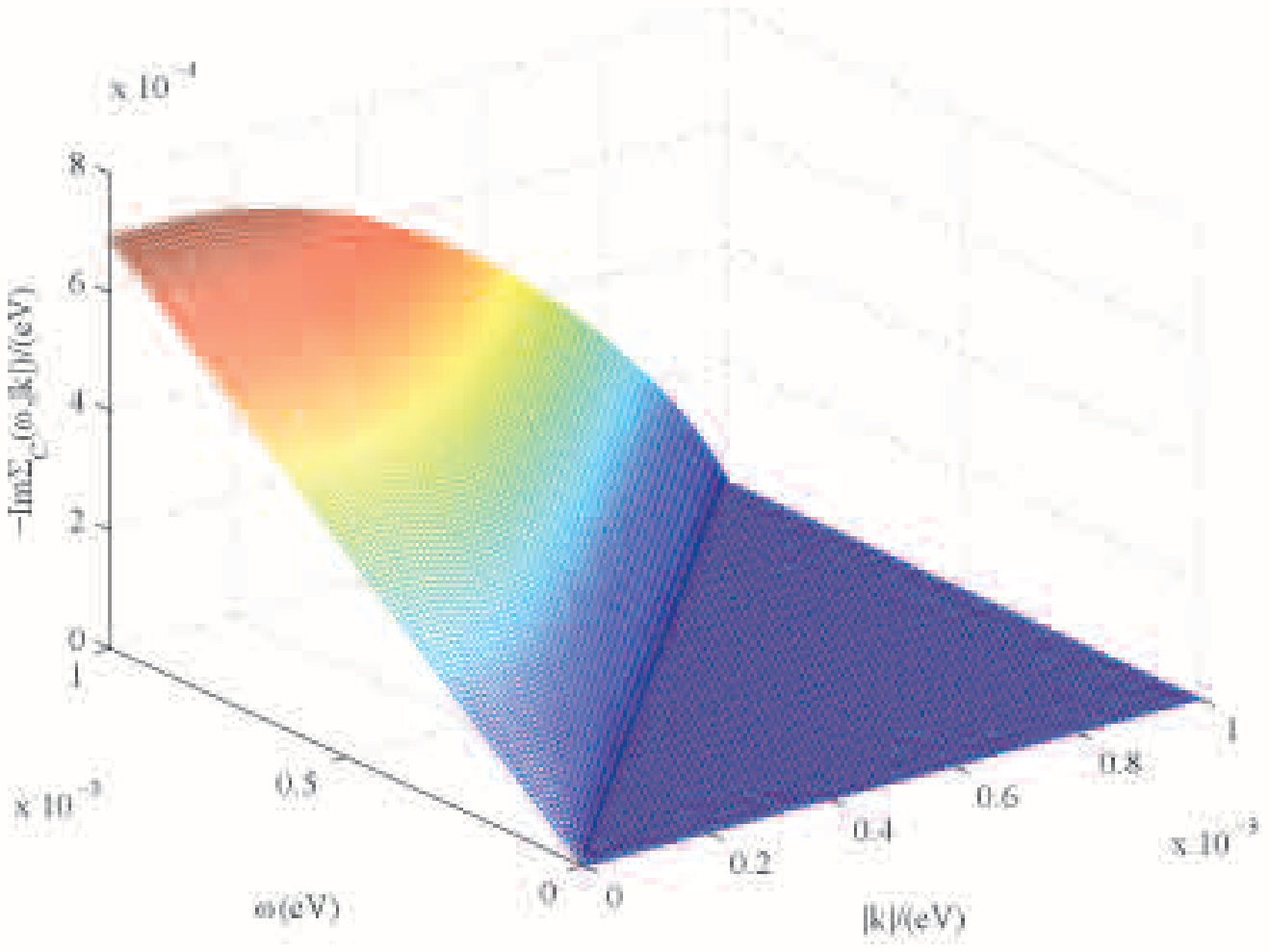}}
    \subfigure[]{
    \label{fig:CoulomP2}
    \includegraphics[width=3.0in]{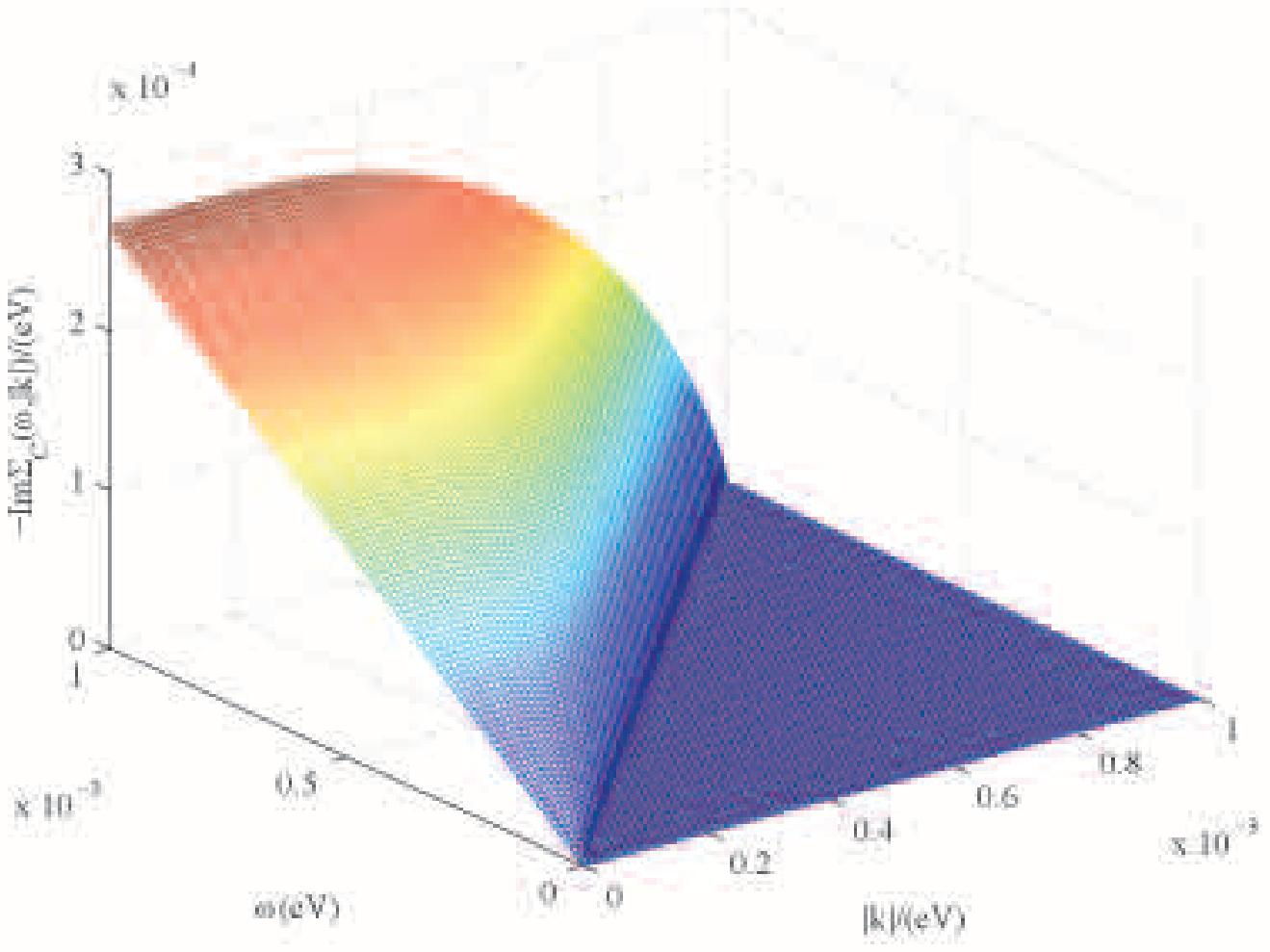}}
     \subfigure[]{
    \label{fig:CoulomP3}
    \includegraphics[width=3.0in]{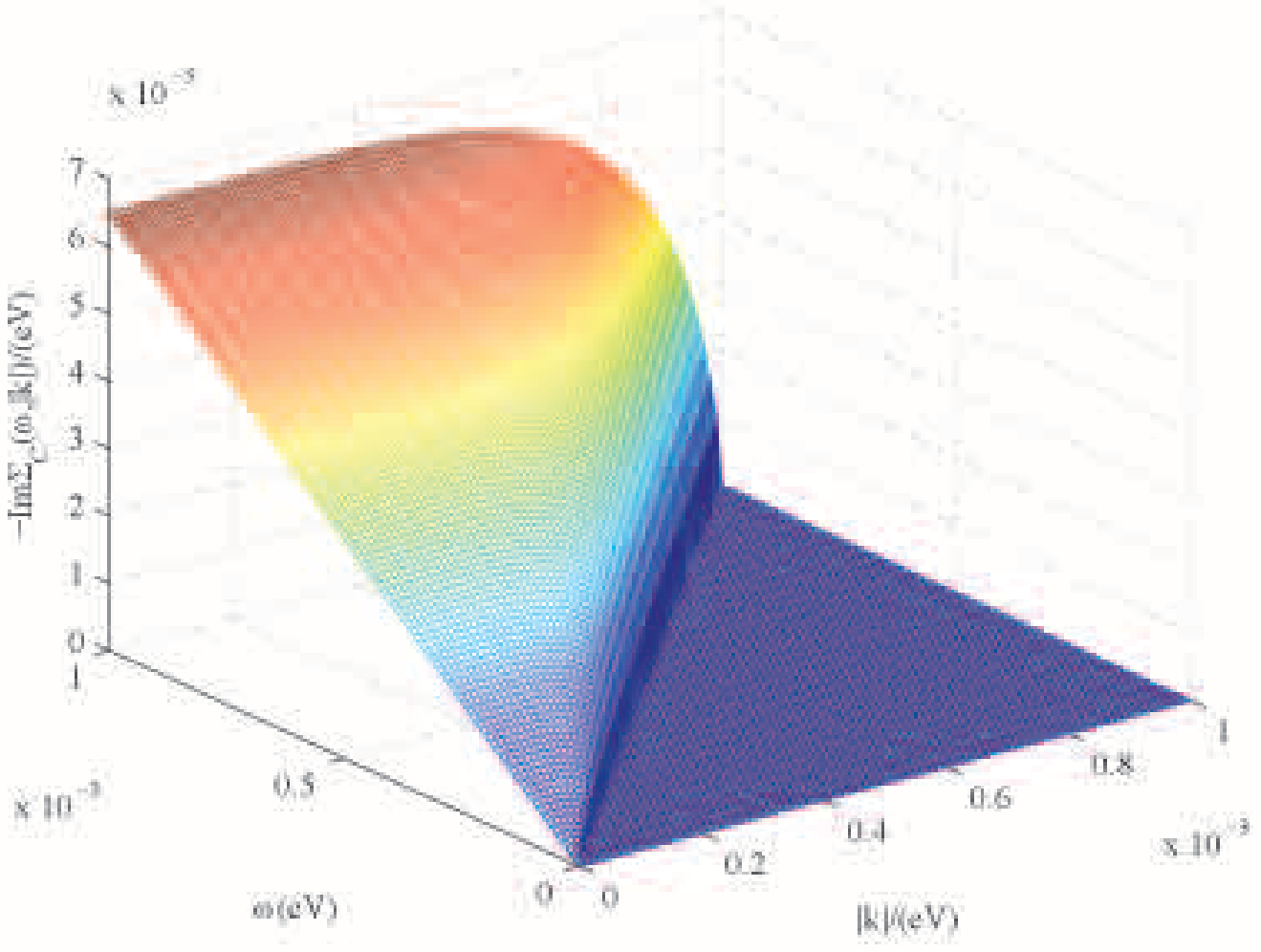}}
    \subfigure[]{
    \label{fig:CoulomP4}
    \includegraphics[width=3.0in]{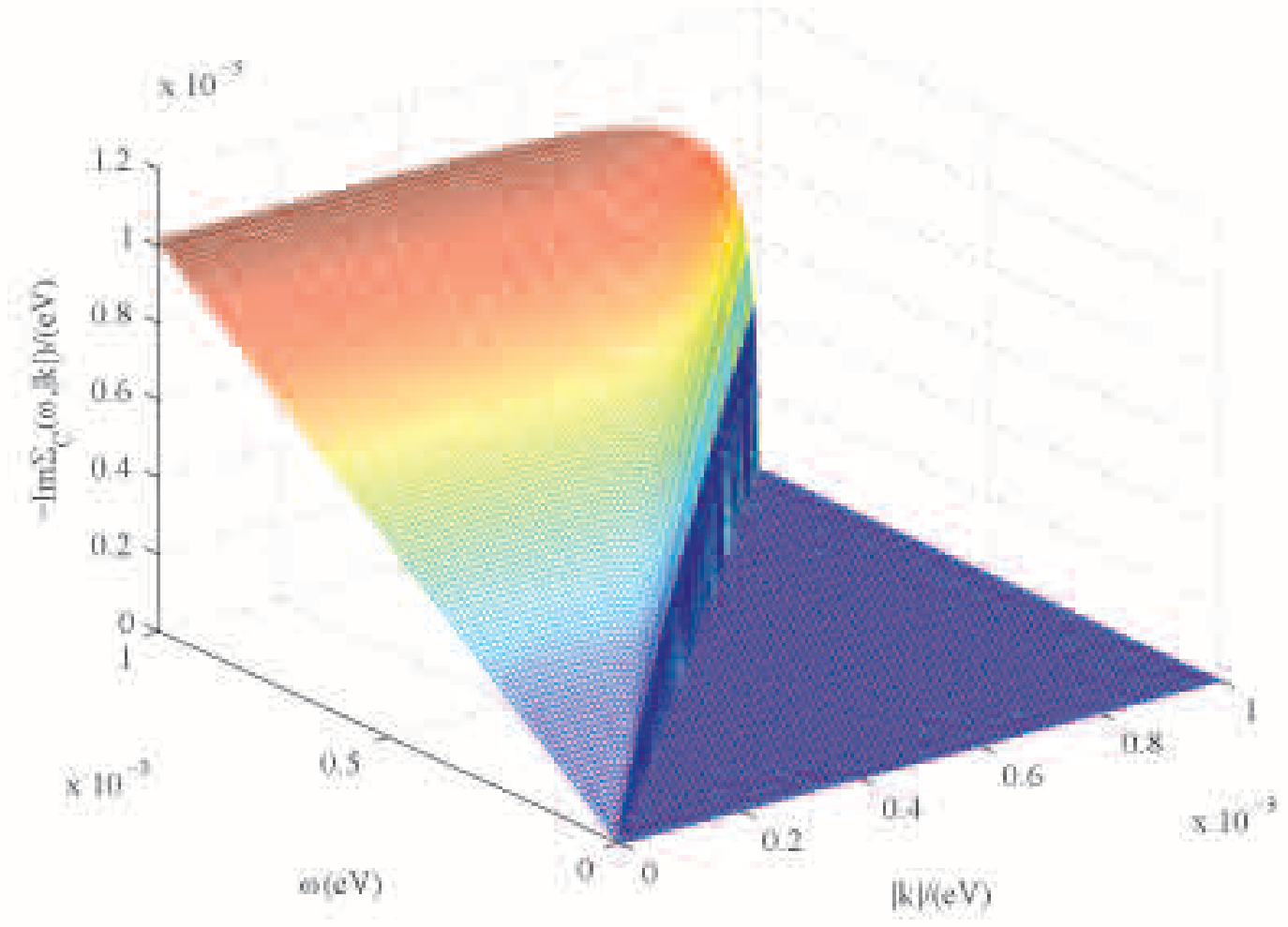}}
 \caption{Damping rate for Dirac fermion due to the Coulomb interaction. $\lambda^{-1}$ are
 $0.001$, $0.01$, $0.1$, and $1$ in (a), (b), (c), (d) respectively.}
     \label{fig:CoulombP}
\end{figure}

The damping rate
$\mathrm{Im}\Sigma_{\mathrm{C}}\left(\omega,\mathbf{k}\right)$ at
general energy-momentum is calculated by numerical methods, with
results shown in Fig. \ref{fig:CoulombP}.

We can also study the self-consistent equations of damping rate and
polarization function. As in the case of gauge field, we consider
only the zero-momentum limit. Now the equation for fermion damping
rate is
\begin{eqnarray}
\mathrm{Im}\Sigma(\omega,T) &=&-\frac{1}{2\pi^2}\int_{0}^{+\infty}
d|\mathbf{q}| \int_{-\infty}^{+\infty} d\omega_{1}
\frac{\mathrm{Im}\Sigma(\omega_{1},T)}
{(\omega_{1}-|\mathbf{q}|)^{2} +
\left(\mathrm{Im}\Sigma(\omega_{1},T)\right)^{2}}
\mathrm{Im}\left[\frac{1}{\frac{|\mathbf{q}|}{\lambda}+D_1
\left(\omega_{1} - \omega,\mathbf{q},T\right)}\right]\nonumber\\
&&\times\left[n_{B}(\omega_{1}-\omega) + n_{F}(\omega_{1})\right],
\end{eqnarray}
where $D_{1}(\varepsilon,\mathbf{q},T)$ was given by
(\ref{eqn:SelfConsistentD1Im}) and (\ref{eqn:SelfConsistentD1Re}).
Due to the special form of the bare interaction function
$|\mathbf{q}|^{-1}$, the energy $\omega$ can be completely scaled
out in the whole set of integral equations. This is not possible for
$|\mathbf{q}|^{-2}$, as discussed in Sec. 3. Moreover, such bare
term ensures the absence of divergence. In the absence of any bare
term, the re-scaling can still be performed, but the equation is
divergent, as discussed in Sec. 2. Therefore, the Coulomb
interaction is very special and turns out to be a good candidate for
producing marginal Fermi liquid behavior.

\begin{figure}[h]
    \subfigure[]{
    \label{fig:CoulombZ}
    \includegraphics[width=3.0in]{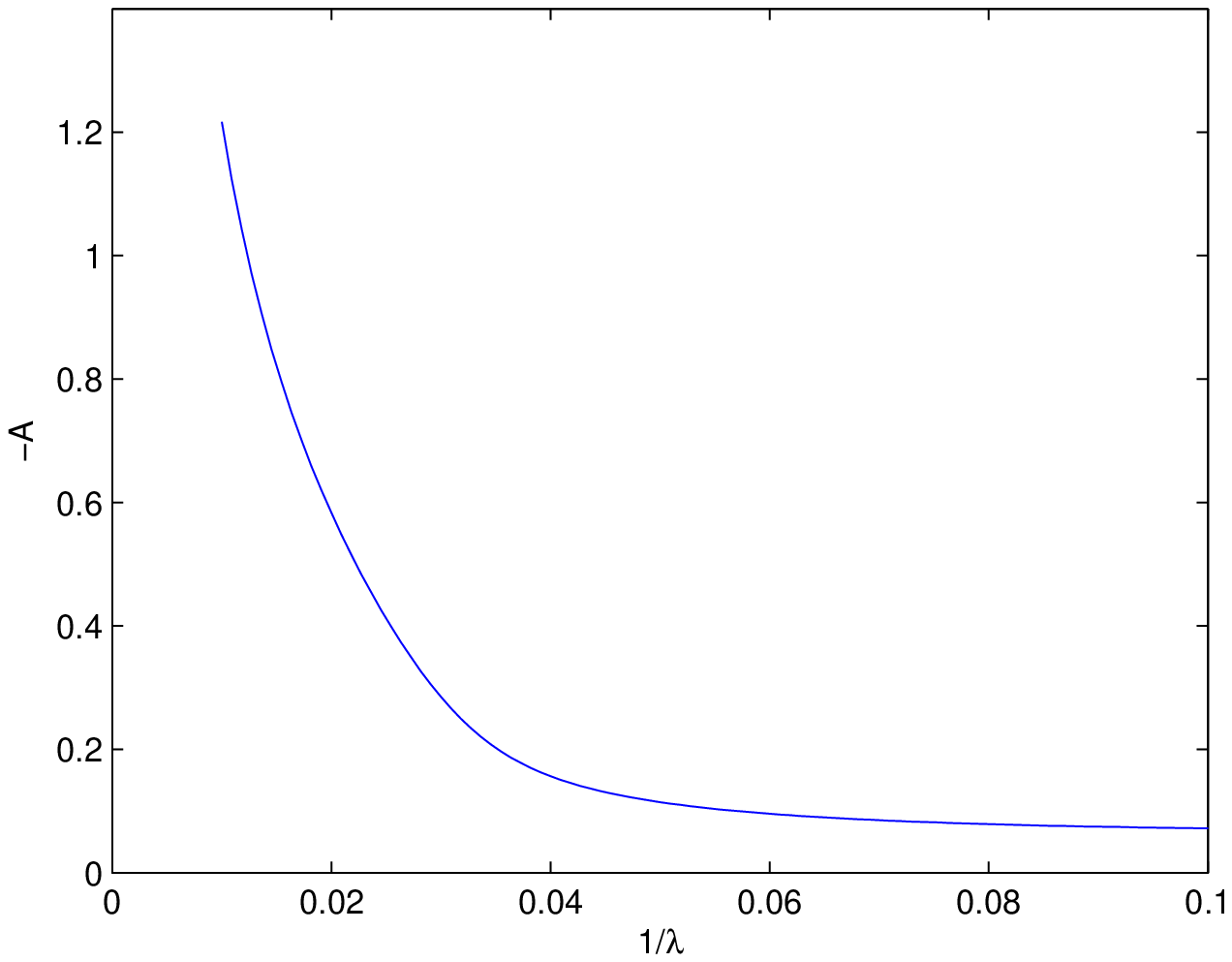}}
    \subfigure[]{
    \label{fig:CoulombT}
    \includegraphics[width=3.0in]{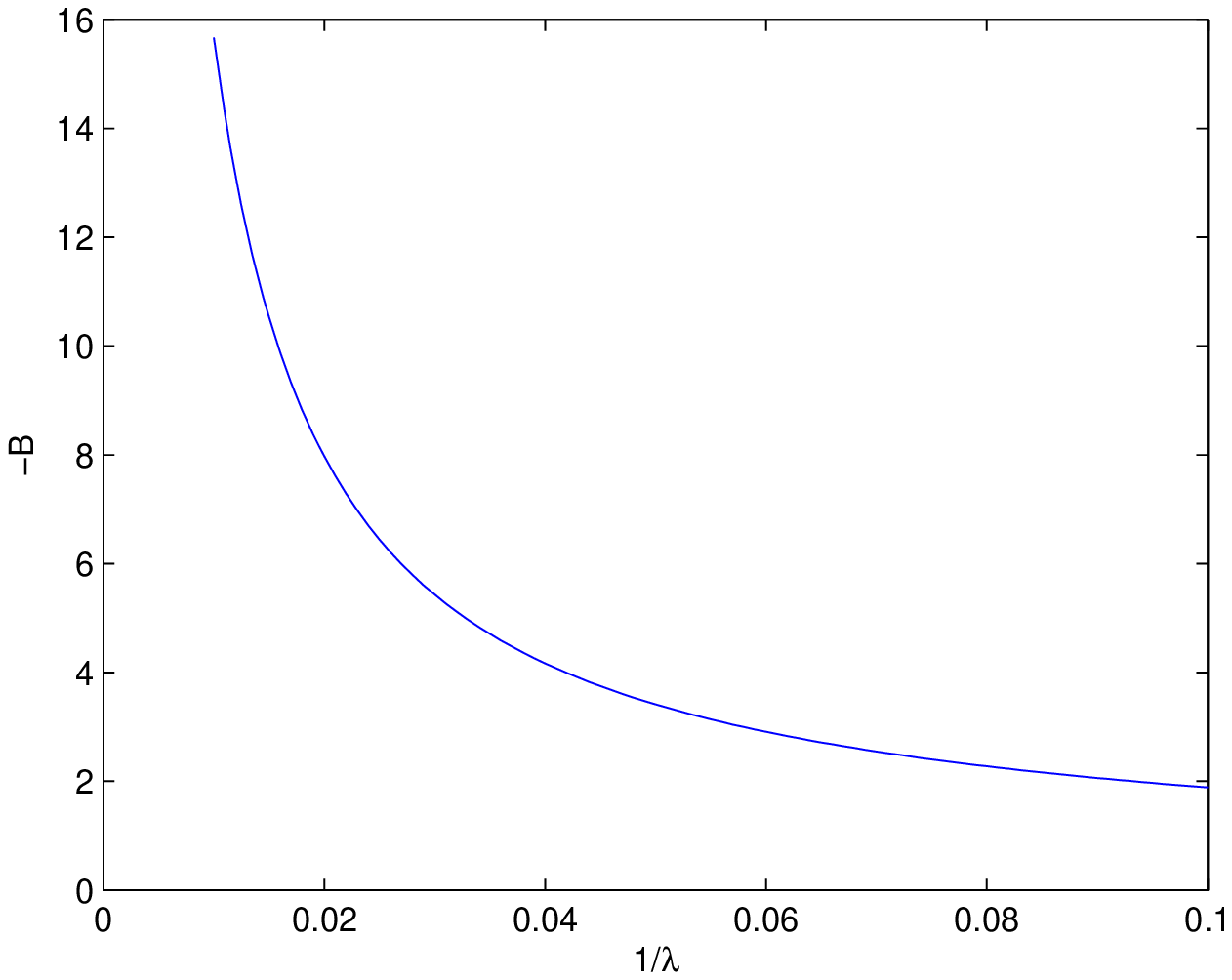}}
 \caption{(a) The relation between $A$ and $1/\lambda$.
 (b) The relation between $B$ and $1/\lambda$.}
\end{figure}

With the help of scaling analysis, it is easy to show that
$\mathrm{Im}\Sigma(\omega) \propto \omega$ in the limit $\omega \gg
T$ and $\mathrm{Im}\Sigma(T) \propto T$ in the limit $\omega \ll T$,
respectively. As before, we define
$\frac{\mathrm{Im}\Sigma(\omega)}{\omega}=A$ and
$\frac{\mathrm{Im}\Sigma(T)}{T}=B$, then their dependence on
$\lambda$ can be obtained by numerical computation. The results are
presented in Fig. \ref{fig:CoulombZ} and Fig. \ref{fig:CoulombT}.

The linear dependence of fermion damping rate on energy/temperature
may appear in other counterpart of the U(1) gauge interaction. It
was discovered in the model of nodal quasiparticles coupled to
critical fluctuation of some superconducting order parameter
\cite{Vojta, Khveshchenko012}. This model, albeit having different
physical contents, share one common feature with Coulomb
interaction: the coupling of massless Dirac fermions to some
singular boson mode. This is in form analogous to U(1) gauge
interaction. However, the calculation of fermion damping rate caused
by gauge interaction meets with divergences. According to our
theoretical and numerical computations, it seems that the kinetic
term for gauge field has to be explicitly included in order to get
meaningful results.

Aji and Varma constructed an interesting dissipative quantum 2D XY
model and showed that it produces marginal Fermi liquid behaviors
\cite{Aji}. Recently, behavior of marginal Fermi liquid type was
also found in a $d$-dimensional field theory with the help of
AdS/CFT correspondence \cite{Faulkner}.

We also note that a linear-in-$T$ quasiparticle damping rate was
already pointed out in two early papers \cite{Ioffe, Lee92}. In both
of these works, the linear-in-$T$ behavior is attributed to the
scattering of quasiparticiles by an emergent U(1) gauge field.
However, such damping rate is that of spinless holons, rather than
fermions. Therefore, the interacting system considered in these
papers, though very interesting, can not be identified as a marginal
"Fermi" liquid.

\section{Summary and Discussion\label{sec:Summary}}

In summary, we present a detailed calculation of the damping rate of
massless Dirac fermions due to U(1) gauge field in QED$_{3}$. When
the theory contains no Maxwell term for gauge field, the fermion
damping rate is found to diverge at both zero and finite
temperatures within perturbation theory. There is still divergence
in the self-consistent equations for fermion damping rate and gauge
boson propagator. Once the Maxwell term for gauge field is included
into the self-consistent equations, the fermion damping rate is
well-defined and display non-Fermi liquid behaviors at both zero and
finite temperatures.

From the first sight, the existence of divergence in fermion damping
rate in QED$_{3}$ theory without Maxwell term might restrict its
validity in understanding high temperature superconductors. However,
in reality this is not as severe as it looks. Such effective theory
applies to both the half-filling state and the underdoped region of
high temperature superconductors. In the half-filling state, the
massless Dirac fermions undergo a pairing instability towards the
chiral symmetry breaking phase. Now the fermionic excitations are
suppressed by the dynamically generated mass gap and thus it is
usually not necessary to study the damping rate. In the underdoped
region, the gauge field couples not only to massless Dirac fermions,
but also to an additional boson field which describes the motion of
charged holons. The interaction between gauge field and holons
contributes a vacuum polarization function to the gauge boson
propagator, which might be able to eliminate the divergence
appearing in the fermion damping rate. To address this issue, it is
essential to carefully study the whole interacting system,
especially the holon-gauge coupling.

At present, we are unable to eliminate the divergence appearing in
the momentum dependence of Dirac fermion damping rate. Such
divergence might also be cured by the self-consistent (Eliashberg)
treatment, but the coupled equations become much more complicated
than the zero-momentum limit and are hence hard to be solved
numerically. We will study this problem further, either by improving
numerical methods or by developing novel theoretical approaches.

When the U(1) gauge theory is used to describe the anomalous
properties of high temperature superconductors, the Dirac fermions
are usually not the physical electrons. They might be fermionic
spinons \cite{Leereview, Kim97, Kim99, Rantner}, fermionic holons
\cite{Kaul}, or topological fermions \cite{Franz, Herbut}, depending
on the physical motivations and the procedures of deriving the
effective field theory. Therefore, the damping rate studied in this
work could be directly compared with experiments only after
including the additional degrees of freedom. However, since the U(1)
gauge interaction of massless Dirac fermions appears naturally in a
number of correlated electron systems, we believe it is interesting
to carefully study the damping rate and other physical quantities of
the "unphysical" Dirac fermions.

\section*{Acknowledgments}

We would like to thank W. Li for helpful discussions. This work is
supported by National Science Foundation of China No. 10674122.

\end{document}